\documentclass[10pt]{IEEEtran}
  
\newcommand{\mbf}{\mathbf}

\usepackage[ruled,linesnumbered]{algorithm2e}
\usepackage{url}
\usepackage{multirow}

\usepackage{graphicx}
\usepackage{amssymb}
\usepackage{amsmath}
\usepackage{amsfonts} 
\usepackage{amsthm}
\usepackage{color}
\usepackage{epstopdf}
\usepackage{verbatim} 
\usepackage[noadjust]{cite}
\usepackage{threeparttable}

\usepackage{soul}
\usepackage{verbatim, url}

\newcounter{theorem}
\newcounter{def}
\newcounter{mylea}
\newcounter{corollary}
\usepackage{fancyhdr}

\newtheorem{lemma}[mylea]{Lemma}

\begin{document}


\title{{Simulation Algorithms with Exponential Integration for Time-Domain Analysis of Large-Scale Power Delivery Networks}}
\author{
Hao~Zhuang, \IEEEmembership{Student Member, IEEE,} 
Wenjian~Yu, \IEEEmembership{Senior Member, IEEE,} 
Shih-Hung~Weng, 
Ilgweon~Kang, \IEEEmembership{Student Member, IEEE,}
Jeng-Hau~Lin, \IEEEmembership{Student Member, IEEE,}   
Xiang Zhang, \IEEEmembership{Member, IEEE,} 
Ryan Coutts,  
and~Chung-Kuan~Cheng, \IEEEmembership{Fellow, IEEE}\\  
\thanks{Manuscript received in May 2015; Revised in August, 2015 and November 2015. 
Accepted in Jan, 2016. } 
\thanks{ This work was supported in part by NSF-CCF 1017864, NSFC 61422402 and NSFC 61274033. 
}
\thanks{H. Zhuang, I.~Kang, J.-H. Lin, and C. K. Cheng are with the Department of Computer Science and Engineering, University of California, San Diego, CA. (e-mail: hao.zhuang@cs.ucsd.edu, igkang@ucsd.edu, jel252@eng.ucsd.edu, ckcheng@ucsd.edu)} 
\thanks{W. Yu is with Tsinghua National Laboratory for Information Science and Technology,  the Department of Computer Science and Technology, Tsinghua University, Beijing, China. (e-mail: yu-wj@tsinghua.edu.cn) } 
\thanks{X. Zhang and R. Coutts are with Department of Electrical and Computer Engineering, University of California, San Diego, CA. (e-mail:xiz110@eng.ucsd.edu,~rmcoutts@eng.ucsd.edu) }
\thanks{S.-H. Weng was with the Department of Computer Science and Engineering, University of California, San Diego, CA. He is now with Facebook, Inc., Menlo Park, CA, USA.
(e-mail: 
shweng@fb.com)
}

 
}
 



\twocolumn  
\maketitle


\begin{abstract}
We design an algorithmic framework using matrix exponentials for time-domain simulation of power delivery network (PDN). 
Our framework can reuse factorized matrices to simulate the large-scale linear PDN system with variable step-sizes. In contrast, current conventional PDN simulation solvers have to use
fixed step-size approach in order to reuse factorized matrices generated by the expensive matrix decomposition.  Based on the proposed exponential integration framework, we design a PDN solver \emph{R-MATEX} with the flexible time-stepping capability. The key operation of matrix exponential and vector product (MEVP) is computed by the rational Krylov subspace method. 

To further improve the runtime, we also propose a distributed computing framework \emph{DR-MATEX}. DR-MATEX reduces Krylov subspace generations caused by frequent breakpoints from a large number of current sources during simulation. By virtue of the superposition property of linear system and scaling invariance property of Krylov subspace, DR-MATEX can divide the whole simulation task into subtasks based on the alignments of breakpoints among those sources. The subtasks are  processed in parallel at different computing nodes without any communication during the computation of transient simulation.  The final  result is obtained by summing up the partial results among all the computing nodes after  they finish the assigned subtasks. {Therefore, our computation model belongs to the category known as \emph{Embarrassingly Parallel}} model. 

Experimental results show R-MATEX and DR-MATEX can achieve up to around  $14.4\times$ and $98.0\times$  runtime speedups over traditional trapezoidal integration based solver with fixed time-step approach.

\end{abstract}

\begin{IEEEkeywords}
Circuit simulation, power delivery/distribution networks, power grid, time-domain simulation, transient simulation, matrix exponential, Krylov subspace method, parallel processing. 
\end{IEEEkeywords} \ 
\IEEEpeerreviewmaketitle

\section{Introduction}
\label{sec:intro} 

\IEEEPARstart{M}{odern} VLSI design verification relies 
heavily on the analysis of power  delivery  network (PDN)
to estimate power supply noises \cite{Najm03_DAC, Nassif00_Fast,Brooks07_DATE,Lin04_CICC,Lin01_ICCAD,Skadron12, Skadron14_ISCA,Skadron14_ASPDAC}. 
The performance of power delivery network highly impacts on the quality of 
global, detailed and mixed-size placement \cite{ljw14_DAC, Pan05_ICCAD, ljw15_TCAD},  
clock tree synthesis \cite{Xiao10_ICCAD}, global and detailed routing \cite{Zhang12_ICCAD}, as well as timing \cite{Kahng13_ICCAD} and power optimization.  
Lowering supply voltages, increasing current densities as well as tight design margins demand more accurate large-scale PDN simulation. 
Advanced technologies \cite{Zhuo12_ICCAD,Zeng10_DAC}, three dimensional (3D) IC structures \cite{Zhuang13_ICCCAS,Lim14_ICCAD,xiang14}, and increasing complexities of system designs all make VLSI PDNs extremely huge and the simulation tasks time-consuming and computationally challenging. 
Due to the enormous size of modern designs and long simulation runtime of many cycles, instead of general nonlinear circuit simulation \cite{Zhuang15_emc, Zhuang15_DAC}, PDN is often modeled as a large-scale linear circuit with voltage supplies and time-varying current sources \cite{Nassif08_Power, zuochang08, Li12_Tau}. Those linear matrices are obtained by parasitic extraction process \cite{Zhuo14_DAC, Zhuang2012, Yu14_TCAD, Yu2013, Lin04_CICC}. 
After those processes, we need time-domain large-scale linear circuit simulation to obtain the transient behavior of PDN with above inputs. 
 
Traditional methods in linear circuit simulation solve differential algebra equations (DAE) 
numerically in explicit ways, e.g., forward Euler (FE), or implicit ways, 
e.g., backward Euler (BE) and trapezoidal (TR), which are all based on low order polynomial approximations for DAEs \cite{Leon75}.
Due to the stiffness of systems, which comes from a wide range of time constants of a circuit, 
the explicit methods require extremely small time step sizes to ensure the stability.
In contrast, implicit methods can handle this problem with relatively large time steps because of their larger stability regions. 
However, at each time step, these methods have to solve a linear system, which is sparse and often ill-conditioned.
Due to the requirement of a robust solution,
compared to iterative methods \cite{Saad03},
direct methods \cite{Davis06} are often favored for VLSI circuit simulation, and thus adopted by state-of-the-art power grid (PG) solvers in TAU PG simulation contest \cite{Yu12, Yang12, Xiong12}. 
Those solvers only require  one matrix factorization (LU or Cholesky factorization) 
at the beginning of the transient simulation.  
Then, at each fixed time step, the following transient computation  requires only pairs of   forward and backward substitutions, which achieves better efficiency over adaptive stepping methods by reusing the factorization matrix \cite{Xiong12, Yu12, Li12_Tau} in their implicit numerical integration framework.  
However, the maximum of step size choice is limited by the smallest distance  $h_{upper}$ among the breakpoints \cite{Nagel75}. Some engineering efforts are spent to break this limitation by sacrificing the accuracy. 
In our work, we always obey the upper limit $h_{upper}$ of time step to maintain the fidelity of model, which means the fixed time step $h$ cannot go beyond   $h_{upper}$ in case of missing breakpoints.


Beyond traditional methods, 
a  class of methods called matrix exponential
time integration has been embraced by MEXP \cite{Weng12_TCAD}.
The major complexity is caused by matrix exponential computations. 
MEXP utilizes standard Krylov subspace method \cite{Saad92} to approximate matrix exponential and vector product. MEXP can solve the DAEs with much higher order polynomial approximations than traditional ones \cite{Weng12_TCAD, Saad92}.
Nevertheless, when simulating stiff circuits with standard Krylov subspace method, it requires the large dimension of subspace in order to preserve the accuracy of MEXP approximation and poses memory bottleneck and degrade the adaptive stepping performance of MEXP.

Nowadays, the emerging multi-core and many-core platforms bring powerful computing resources and opportunities for parallel computing. Even more, cloud computing techniques \cite{cloud} drive distributed systems scaling to thousands of computing nodes 
\cite{mapreduce, xiang09, spark10}, etc. 
Distributed computing systems have been incorporated into products of many leading EDA companies and in-house simulators \cite{He14_emc, Jia15_TCAD, Wang13, Tan13_ICCAD, Feng08}. 
However, building scalable and efficient distributed algorithmic framework for transient linear circuit simulation is still a challenge to leverage these powerful computing tools.    
The papers \cite{Zhuang14_DAC, gander2013paraexp} show  great potentials by parallelizing matrix exponential based method to achieve the runtime performance improvement and maintain high accuracy. 


In this work, we develop a transient simulation framework  
using matrix exponential integration scheme, \emph{MATEX}, for  PDN simulation.
Following are the challenges we need to address. 
First,   when the circuit is stiff, the standard Krylov subspace has convergence problem and slows down the computation of MEVP.
Second, the frequent time breakpoints due to the transitions of PDN current sources modeling  triggers the generations of Krylov subspace. 
Therefore, we might gain performance where we leverage the large time stepping, but we also lose runtime for the small step size.
Our contributions are listed as below:
\begin{itemize} 
\item   MEVP in MATEX is efficiently computed by rational or invert Krylov subspace method. 
Compared to the commonly adopted framework using TR  with fixed time step (TR-FTS), the proposed MATEX can reuse factorized matrix at the beginning of transient simulation to perform flexible adaptive  time stepping. 
\item Among different Krylov subspace methods, we find rational Krylov subspace is the best strategy for MEVP in PDN simulation. Therefore, we design R-MATEX based on that and achieve  {
up to around $15\times$} runtime speedup  against the benchmarks over the traditional method  TR-FTS with good accuracy. 
\item Furthermore, DR-MATEX is designed to improve R-MATEX with distributed computing resources.
\begin{itemize}
\item First, PDN's current sources are partitioned into groups based on their alignments. 
They are assigned to different computing nodes.
Each node runs its corresponding PDN transient simulation task and has no communication overhead with other nodes.
\item
After all nodes finish the simulation computations, the results are summed up based on the linear superposition property of the PDN system.
\item Proposed current source partition can reduce the chances of generating Krylov subspaces and prolong the time periods of reusing computed subspace at each node, which brings huge computational advantage and achieves up to $98\times$ speedup over traditional method TR-FTS. 
\end{itemize}
\end{itemize}  
The rest of this paper is organized as follows. Sec.  \ref{sec:method} introduces the background of
linear circuit simulation and matrix exponential formulations.
Sec. \ref{sec:advanced} illustrates the Krylov techniques to accelerate matrix exponential and vector product computation.   
Sec. \ref{sec:matex} presents MATEX circuit solver and the parallel framework DR-MATEX.
Sec. \ref{sec:exp_results} shows numerical results 
and Sec. \ref{sec:conclusion} concludes this paper.

\section{Background}
\label{sec:method}
\subsection{Transient Simulation of Linear Circuit}
Transient  simulation of linear circuit is the foundation of modern PDN simulation. It is formulated as DAEs via modified nodal analysis (MNA),
\begin{eqnarray}
    \label{eqn:dae} 
    \mbf{C}\dot{\mbf{x}}(t) = - \mbf{G}\mbf{x}(t) + \mbf{B}\mbf{u}(t),
\end{eqnarray}
where  $\mbf{C}$ is the  matrix for  capacitive and inductive elements. $\mbf{G}$ is the matrix for conductance and resistance, and $\mbf{B}$ is the input selector matrix.  $\mbf{x}(t)$ is the vector of time-varying node voltages
and branch currents. $\mbf{u}(t)$  is the vector of  supply voltage and current sources. In PDN, such current sources are often characterized as pulse or piecewise-linear inputs \cite{Nassif08_Power, Li12_Tau} to represent the activities under the networks. To solve Eq. (\ref{eqn:dae}) numerically, the system is discretized with time step $h$ and transformed to a linear algebraic system. Given an initial condition $\mbf x(0)$ from DC analysis or previous time step $\mbf x(t)$ and a time step $h$, $\mbf x(t+h)$ can be obtained by traditional \emph{low order approximation} methods \cite{Leon75}. 

\subsection{Traditional Low Order Time Integration Schemes}
\subsubsection{BE} Backward Euler based time integration scheme (Eq.(\ref{eqn:be})) is a robust implicit first-order method. 
\begin{eqnarray}
    \label{eqn:be}
    \Bigg(\frac{\mbf{C}}{h} + 
     {\mbf G} \Bigg) {\mbf{x}}(t+h) 
    =  \frac{\mbf C}{h} 
    \mbf{x}(t) + \mbf{B}     \mbf{u}(t+h)  .
\end{eqnarray}
\subsubsection{TR} Trapezoidal based time integration scheme (Eq.(\ref{eqn:trap})) is a popular implicit second-order method.
\begin{eqnarray}
  \label{eqn:trap}
    \Bigg(\frac{\mbf{C}}{h} +  
    \frac{\mbf G}{2}\Bigg) {\mbf{x}}(t+h)    & = &  
    \Bigg(\frac{\mbf C}{h}- \frac{\mbf G}{2} \Bigg)
    \mbf{x}(t)  \\ &+& \mbf{B} \frac{  \mbf{u}(t) + \mbf{u}(t+h)} {2}. \nonumber
\end{eqnarray}
It is probably the most commonly used strategy for large-scale circuit simulation, which has higher accuracy than BE.

\subsubsection{BE-FTS and TR-FTS} 
\begin{figure}[ht]
	\centering
	\includegraphics[trim = 3.5cm 4cm 5cm 0cm, clip,keepaspectratio, width=0.45\textwidth]{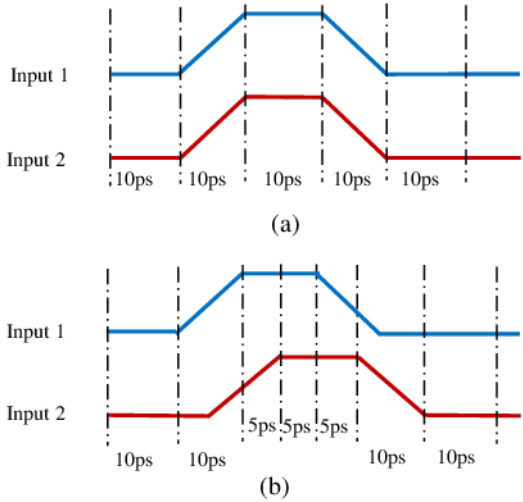}
	\caption{
		Example: Interleave  two input sources to create smaller transition time. ({  a}) Before interleaving, the smallest transition time of the input sources is $h_{upper}=10ps$; { (b)} After interleaving, the smallest transition time of the input sources is $h_{upper}=5ps$. 
	}
	\label{fig:interleave}	
\end{figure}

Methods BE and TR with fixed time step (FTS) $h$ are efficient approaches, which were adopted by the top PG solvers in 2012 TAU PG simulation contest \cite{Xiong12,Yu12,Yang12,Li12_Tau}. 
If only one $h$ is used for the entire simulation, the choice is limited by the minimum breakpoint \cite{Nagel75} distance $h_{upper}$ among all the input sources. 
Fig. \ref{fig:interleave} (a) has $10ps$ as the upper limit for  $h$ in BE-FTS and TR-FTS. When the alignments of inputs change (as shown in Fig. \ref{fig:interleave} (b)) shift by $5ps$, the resulting upper limit for $h$ becomes $5ps$ for the approaches with fixed step size. 
If $h$ is larger than the limit, it is impossible to guarantee the accuracy since we may skip pivot points of the inputs. 

\subsection{Matrix Exponential Time Integration Scheme}

The solution of Eq. (\ref{eqn:dae}) can be obtained analytically \cite{Leon75}. 
For a simple illustration, we convert Eq. (\ref{eqn:dae}) into 
\begin{eqnarray} 
\label{eqn:ode}
    \dot{\mbf{x}}(t) = \mbf{A}\mbf{x}(t) + \mbf{b}(t),
\end{eqnarray}
when $\mbf{C}$ is not singular\footnote{The assumption is to simplify the explanation in this section. After Sec. \ref{sec:imatex}, we use I-MATEX, R-MATEX and DR-MATEX to compute the solution of DAE without inversion of $\mbf C$. Therefore, the methods are suitable for general DAE system, i.e., Eq. (\ref{eqn:dae}) without the assumption here.}, 
$$\mbf{A} = - \mbf{C}^{-1}\mbf{G}~\text{, and}~\mbf{b}(t) = \mbf{C}^{-1}\mbf{B}\mbf{u}(t).$$
Given  a solution at time 
$t$ and a time step $h$, the solution at
$t + h$ is
\begin{eqnarray}  
    \label{eqn:discrete_sol}
    \mbf{x}(t+h) = e^{h\mbf{A}}\mbf{x}(t) + 
    \int^{h}_{0}e^{(h-\tau)\mbf{A}}\mbf{b}(t+\tau)d\tau.
\end{eqnarray}

Assuming that the input $\mbf{u}(t)$ is a piecewise linear (PWL) function of $t$, 
we can integrate the last term of Eq. (\ref{eqn:discrete_sol})  analytically, turning the solution with matrix exponential operator:
\begin{eqnarray} 
\label{eqn:pwl_exact_sol}
    \mbf{x}(t+h)  \nonumber = 
      -\left(\mbf A^{-1}\mbf b(t+h) + \mbf A^{-2} \frac{\mbf b(t+h)- \mbf b(t)}{h}\right) +
        \\
       e^{h\mbf{A}}\left(\mbf{x}(t) + \mbf A^{-1}\mbf b(t) + \mbf A^{-2} \frac{\mbf b(t+h)- \mbf b(t)}{h}\right )  .
\end{eqnarray}
For the time step choice,   breakpoints  (also known as input transition spots (TS) \cite{Zhuang14_DAC}) are the time points where slopes of input  vector change. 
Therefore, for Eq. (\ref{eqn:pwl_exact_sol}), the maximum time step starting from $t$ is $(t_s-t)$,
where $t_s$ is the smallest one in $TS$ larger than $t$. 
In matrix exponential based framework, the limitation of time step size is not the local truncation error (LTE), but the activities among all input sources.



\section{Krylov Subspace Methods for Matrix Exponential and Vector Product (MEVP)}
\label{sec:advanced}
 In PDN simulation, $\mbf A$  is usually above millions and makes the direct computation of matrix exponential $e^{\mbf A}$ infeasible. The alternative way to compute the product is through Krylov subspace method \cite{Saad92}. In this section, we first introduce the background of standard Krylov subspace for MEVP. Then, we discuss invert (I-MATEX) and rational Krylov subspace (R-MATEX) methods, which highly improve the runtime performance for MEVP. 

\subsection{MEXP: MEVP Computation via Standard Krylov Subspace Method}
\label{sec:std}
The complexity of $e^{\mbf A}\mbf v$ can be reduced 
using Krylov subspace method and still maintained  
in a high order polynomial approximation \cite{Saad92}.
MEXP \cite{Weng12_TCAD} uses standard Krylov subspace, 
which uses $\mbf A$ directly to generate subspace basis through Arnoldi process (Algorithm  \ref{algo:arnoldi}). 
First,
we reformulate Eq. (\ref{eqn:pwl_exact_sol}) into
\begin{eqnarray}
	\label{eqn:new_exact}
	\mbf{x}(t+h)
	=
	e^{h\mbf A}(\mbf x(t) + \mbf F(t,h) ) - \mbf P (t,h),
\end{eqnarray}
where 
\begin{eqnarray}
	\mbf F(t,h) =  \mbf A^{-1}\mbf b(t) + \mbf A^{-2} \frac{\mbf b(t+h)- \mbf b(t)}{h}
\end{eqnarray} and 
\begin{eqnarray}
	\mbf P(t,h) =   \mbf A^{-1}\mbf b(t+h) + \mbf A^{-2} \frac{\mbf b(t+h)- \mbf b(t)}{h}  .
\end{eqnarray}
The standard Krylov subspace 
\begin{eqnarray}
	{\rm K_m}(\mbf A, \mbf v) := \text{span}
	{\{ \mbf v, \mbf A\mbf v,
		\cdots, \mbf A^{m-1} \mbf v}\}
\end{eqnarray}
obtained by Arnoldi process
has the relation
\begin{eqnarray}
	\mbf{A}\mbf{ V_m} = 
\mbf {   V_m} \mbf {  H_m}
	+ {  h_{m+1,m} }
	\mbf {  v}_{m+1}\mbf e_m^\mathsf{T}, 
\end{eqnarray}
{ where 
$\mbf {H_m}$ is the upper Hessenberg matrix 
\begin{eqnarray}
\mbf {H_m } = \begin{pmatrix}
 h_{1,1} &   h_{1,2} & \cdots &   h_{1,m-1} &   h_{1,m} \\
 h_{2,1} &   h_{2,2} & \cdots &   h_{2,m-1} &   h_{2,m} \\
0 &   h_{3,2} & \cdots &  h_{3,m-1} &   h_{3,m} \\
\cdots & \cdots & \cdots & \cdots  & \cdots \\
0 & 0  & \cdots &  h_{m,m-1} &   h_{m,m} \\
\end{pmatrix},
\label{eq:hess}
\end{eqnarray}
$\mbf { V_m}$ is a $n \times m $ matrix by $(\mbf v_1, \mbf v_2, \cdots, \mbf v_m)$, and $\mbf e_m$ is the $m$-th unit vector with dimension $n\times 1$.
MEVP is computed via 
\begin{eqnarray}
	e^{h\mbf A} \mbf v \approx \beta \mbf {  V_m} e^{h\mbf {  H}_m} \mbf e_1 .
\end{eqnarray}
The posterior error term is  
\begin{eqnarray}
	\label{eqn:std_error}
   r_m(h)  =  \lVert \beta { h_{m+1,m}} \mbf v_{m+1}
  	\mbf e^{\mathsf{T}} e^{h\mbf{  H}_m} \mbf e_1 \rVert,
\end{eqnarray}
where $\beta = \lVert  \mbf v \rVert $.
However, for an autonomous system $ \mbf C \dot {\mbf x}(t) = - \mbf G \mbf x(t)$  in circuit simulation, we consider the residual between  
$\mbf C \dot {\mbf x}(t) $ and $-\mbf G \mbf  x(t)$, 
which is 
$$ \mbf C \dot {\mbf x}(t) + \mbf G \mbf  x(t) ,$$
instead of  $$ \dot {\mbf x}(t) - \mbf A \mbf  x(t)$$  in
$ \dot {\mbf x}(t) = \mbf A \mbf x(t)$.
This leads to 
\begin{eqnarray}
	\label{eqn:std_err}
	   r(m,h)  =    \lVert\beta  h_{m+1,m}  \mbf C  \mbf 
v_{m+1}	\mbf e^{\mathsf{T}} e^{h\mbf{  H}_m} \mbf e_1 \rVert
\end{eqnarray}
and helps us mitigate the overestimation of the error bound. 

To generate $\mbf x(t+h)$ by Algorithm \ref{algo:arnoldi}, 
we use 
\begin{eqnarray}
\label{eq:lu}
[\mbf L, ~\mbf U] = \text{LU\_Decompose}(\mbf X_1),
\end{eqnarray}
where $$\mbf X_1 = \mbf C ~ \text{and} ~ \mbf X_2 = \mbf G$$ as inputs for standard Krylov subspace.
The error budget $\epsilon$ and Eq. (\ref{eqn:std_error}) are used to determine the convergence when  time step is $h$ and Krylov subspace dimension is $j$  (from line \ref{alg1:conv1} to line \ref{alg1:conv2} in Algorithm \ref{algo:arnoldi}). 
\begin{algorithm}
	\label{algo:arnoldi}
	\caption{MATEX\_Arnoldi}
	\KwIn{ $\mbf L, \mbf U, \mbf X_2, h ,t, \mbf x(t),  \epsilon, \mbf P (t,h),\mbf F (t,h)$  }
	\KwOut{ $\mbf x(t+h), \mbf V_m, \mbf H, \mbf v$}
	{   
		$\mbf v = \mbf x(t) + \mbf F(t,h)$\;
		$\mbf v_1 = \frac{\mbf v }{\lVert \mbf v \rVert}$\;
		\For  {$j=1:m$}
		{
			\label{arnoldi_orth} 
			$\mbf w = \mbf U \backslash  (\mbf L \backslash (\mbf X_2 \mbf v_{j}))$
			{\tcc*[l]{a pair of forward and backward substitutions.}}
			\For {$i = 1:j$}
			{
				$h_{i,j} = \mbf w^T\mbf v_{i}$\; 
				$\mbf w = \mbf w - h_{i,j} \mbf v_{i}$\;
			}
			$h_{j+1,j} = \lVert \mbf w \rVert $\; 
			$\mbf v_{j+1} = \frac{\mbf w }{h_{j+1,j}} $\;
			\If {$r(j,h)  < \epsilon $\label{alg1:conv1}} 
			{
				$m = j$\;
				break\; 
			}\label{alg1:conv2}
		}
		$\mbf x(t+h) = \lVert \mbf v \rVert \mbf {V_m} e^{h \mbf H} \mbf e_1 - \mbf P(t,h)$\;
		\label{alg1:h}
	}
\end{algorithm}

\label{sec:discuss}

The standard Krylov subspace may not be efficient when simulating stiff circuits \cite{Weng12_TCAD, Weng12_ICCAD}.
For the accuracy of approximation of $e^{\mbf A}\mbf v$, a large dimension of Krylov subspace basis is required, which not only brings the computational complexity but also consumes huge amount of memory.
The reason is that the Hessenberg matrix $\mbf H_{m}$ of standard Krylov subspace tends to approximate the large magnitude eigenvalues of $\mbf A$ \cite{Van06}. 
Due to the exponential decay of higher order terms in Taylor's expansion, such components are not the crux of circuit system's behavior 
\cite{Botchev12, Van06}.
Therefore, to simulate stiff circuit, we need to gather more vectors into subspace basis 
and increase the size of $\mbf H_{m}$ to fetch more useful components, which results in both
memory overhead and computational complexity to Krylov subspace generations for each time step.
In the following subsections, we adopt  ideas from \emph{spectral transformation} \cite{Botchev12, Van06} to effectively capture small magnitude eigenvalues in $\mbf A$, leading to a fast and accurate MEVP computation. 

\subsection{I-MATEX: MEVP Computation via Invert Krylov Subspace Method}
\label{sec:imatex}
Instead of $\mbf A$, we use $\mbf A^{-1}$ (or $\mbf G^{-1} \mbf C$) as our target matrix to form
\begin{eqnarray}
	{\rm K_m}(\mbf A^{-1} , \mbf v) := \text{span}
	{\{ \mbf v, \mbf A^{-1}\mbf v, \cdots, \mbf A^{-(m-1)} \mbf v}\}.
\end{eqnarray}
Intuitively, by inverting $\mbf A$, the small magnitude eigenvalues become the large ones
of $\mbf A^{-1}$.
The resulting $\mbf H_{m}$ is likely to capture these
eigenvalues first. Based on Arnoldi algorithm, the invert Krylov subspace has the relation 
\begin{eqnarray}
	\mbf{A}^{-1}\mbf{V_m} =  
	\mbf V_m \mbf H_m + 
	h_{m+1,m} \mbf v_{m+1}\mbf e_m^\mathsf{T}. 
\end{eqnarray}
The matrix exponential $e^{\mbf A} \mbf v$ is calculated as 
\begin{eqnarray}
	e^{\mbf A} \mbf v \approx \beta \mbf V_m e^{h\mbf {H}^{-1}_m} \mbf e_1.
\end{eqnarray}
To put this method into Algorithm 1 is just by modifying the inputs 
$\mbf X_1 = \mbf G$ for the LU decomposition in Eq. (\ref{eq:lu}),
and $\mbf X_2 = \mbf C$.  
In the line \ref{alg1:h} of Algorithm \ref{algo:arnoldi}, 
$$\mbf H = \mbf {H}^{-1}_m$$ for the invert Krylov version.
The posterior error approximation \cite{Zhuang14_DAC} is
\begin{eqnarray}
	\label{eq:err_inverted_krylov}
	    r_m(h) 
	= 
	\lVert \beta  h_{m+1,m}  \mbf A  \mbf v_{m+1} 
	\mbf e^\mathsf{T}_m {\mbf H}^{-1}_{m}
	e^{h {\mbf H}}
	\mbf e_1 \rVert,
\end{eqnarray}
which is derived from residual based error approximation in
\cite{Botchev12}.
However, as mentioned in Sec. \ref{sec:std}, we consider the residual of $\left(\mbf C \dot {\mbf x}(t) + \mbf G \mbf x(t) \right)$, 
instead of $\left( \dot {\mbf x}(t) - \mbf A \mbf  x(t) \right)$, 
which leads to 
\begin{eqnarray}
	\label{eq:er_inverted_krylov}
   r(m,h) 
	= 
	\lVert   \beta h_{m+1,m} \mbf G   \mbf v_{m+1} 
	\mbf e^\mathsf{T}_m {\mbf H}^{-1}_{m}
	e^{h {\mbf H}_{m}}
	\mbf e_1  \rVert .
\end{eqnarray}
{We use Eq. (\ref{eq:er_inverted_krylov}) for the line \ref{alg1:conv1} of Alg. \ref{algo:arnoldi}.

\subsection{R-MATEX: MEVP Computation via Rational Krylov Subspace Method}
\label{krylov_rat_err}
The shift-and-invert Krylov subspace basis 
\cite{Van06} 
is designed
to confine the spectrum of
$\mbf A$.
Then, we generate Krylov subspace via
\begin{eqnarray}
	& &  \mbf {K_m} ((\mbf I- \gamma \mbf {A} )^{-1}, \mbf v) :=  \\
	& &  \text{span} \{ \mbf v, (\mbf I- \gamma\mbf { A} )^{-1}\mbf v,
	\cdots,
	(  \mbf I -  \gamma \mbf{A})^{-(m-1)} \mbf v \}, \nonumber 
\end{eqnarray}
where $\gamma$ is a predefined parameter.
With this shift, all the eigenvalues' magnitudes are larger than one. 
Then the inverse limits the magnitudes smaller than one.
According to \cite{Botchev12, Van06},
the shift-and-invert basis for matrix exponential based transient simulation
is not very sensitive to $\gamma$, once it is set to around the order near time steps 
used in transient simulation.
The similar idea has been applied to simple power grid simulation 
with matrix exponential method
\cite{Zhuang13_ASICON}.
Here, we generalize this technique and integrate into MATEX.
The Arnoldi process constructs $\mbf{V_m}$ and $\mbf{H_m}$. 
We have 
\begin{eqnarray}
	(\mbf{I}-  \gamma \mbf {A})^{-1} \mbf{ V}_m =
	\mbf{ V}_m\mbf{ {H}}_m +  {h}_{m+1,m} \mbf { v}_{m+1} \mbf e^\mathsf{T}_m . 
\end{eqnarray}
We can project the $e^{\mbf{{A}}}$ onto the rational Krylov subspace as follows.
\begin{eqnarray}
	\label{eq:scale}
	e^{\mbf{A}h} \mbf{v}  \approx
	 \beta \mbf{  V}_m e^{h\frac{\mbf I - \mbf { {H}}_m^{-1}}{\gamma}}\mbf e_1.
\end{eqnarray}
In the line \ref{alg1:h} of Algorithm \ref{algo:arnoldi},  
$${\mbf{H}} = \frac{\mbf I - \mbf { {H}}_m^{-1}}{\gamma}.$$  
Following the same procedure \cite{Zhuang14_DAC, Botchev12}, 
the posterior error approximation is derived as  
\begin{eqnarray}
	\label{eq:err_rational_krylov}
   r_m(h) 
	= 	\lVert  \beta
	{h}_{m+1,m}  \frac{\mbf I - \gamma \mbf A_m}{\gamma}
	\mbf {  v}_{m+1} 
	\mbf e^\mathsf{T}_m \mbf{  H}_m^{-1}  
	e^{ h\mbf{H}} 
	\mbf e_1  \rVert .
\end{eqnarray}
Note that in practice, instead of 
computing $(\mbf I-\gamma \mbf{A})^{-1}$ directly,
$(\mbf C + \gamma \mbf G)^{-1}\mbf C $ is utilized. 
The corresponding Arnoldi process shares  
the same skeleton of Algorithm \ref{algo:arnoldi} 
with input matrices 
$$\mbf X_1 = (\mbf C+ \gamma \mbf G)$$ for the LU decomposition Eq. (\ref{eq:lu}),
and $$\mbf X_2 = \mbf {C}.$$
The residual estimation is 
\begin{eqnarray}
	\label{eq:er_rational_krylov}
  r(m,h)  
	= 	\lVert     \beta
	{h}_{m+1,m}
	\frac{\mbf C  + \gamma \mbf G}{\gamma} 
	\mbf {  v}_{m+1} 
	\mbf e^\mathsf{T}_m \mbf{  {H}}_m^{-1}  
	e^{ h\mbf{  H}_m} 
	\mbf e_1 \rVert .
\end{eqnarray}
Then, we plug Eq. (\ref{eq:er_rational_krylov}) into the line \ref{alg1:conv1} of Algorithm \ref{algo:arnoldi}.
}

\subsection{Regularization-Free MEVP Computation}
When $\mbf C$ is a singular matrix, MEXP \cite{Weng12_TCAD} needs the regularization process
\cite{Chen12_TCAD} to remove the singularity of DAE in Eq. (\ref{eqn:dae}).
It is because MEXP needs factorize $\mbf C$ directly to form the input $\mbf X_1$ for Algorithm 1. 
This brings extra computational overhead when the case is large \cite{Chen12_TCAD}. 
It is not necessary if we can obtain the generalized eigenvalues and corresponding eigenvectors for matrix pencil $(-\mbf G, \mbf C)$. 
Based on \cite{Wilk:79}, we derive the following lemma,
\begin{lemma}
	Considering a homogeneous system 
	$$ \mbf C \mbf {  \dot{x}} = -\mbf G \mbf x.$$ 
	$\mbf u$ and $\lambda$ are the eigenvector and eigenvalue 
	of matrix pencil $(-\mbf G, \mbf C)$, then $$\mbf x =  e^{t\lambda} \mbf u$$ is a solution of the system.
\end{lemma}
{ 
\begin{proof}\footnote{We repeat the proof from \cite{Wilk:79} with some modifications for our formulation.} 
If $\lambda$ and $\mbf u$ are an eigenvalue and eigenvector of a generalized eigenvalue problem $$-\mbf G \mbf u = \lambda \mbf C \mbf u.$$
Then, $ \mbf x = e^{t\lambda} \mbf u$ is the solution of $\mbf C \mbf {\dot x} =  - \mbf G \mbf x$. 
\end{proof}}
\vspace{+0.07in}
Because we do not need to compute $\mbf C^{-1}$ explicitly during Krylov subspace generation, I-MATEX and R-MATEX are regularization-free. Instead, we factorize $\mbf G$ for invert Krylov subspace basis generation (I-MATEX), 
or $(\mbf C+\gamma\mbf G)$ for rational Krylov subspace basis (R-MATEX).\footnote{It is also applied to the later work of DR-MATEX in Sec. \ref{sec:dist_frame}.}}
Besides, their Hessenberg matrices Eq. (\ref{eq:hess}) are  invertible, 
which contain corresponding important generalized eigenvalues/eigenvectors from 
matrix pencil $(-\mbf G, \mbf C)$, and define the behavior of linear dynamic system in Eq. (\ref{eqn:dae}) of  interest. 
  
\subsection{Comparisons among Different Krylov Subspace Algorithms for MEVP Computation}
\label{sec:compare_krylov}
 
\begin{figure}[t]
	\centering
	\includegraphics[trim = 0.7cm 0.0cm 0.0cm 0.0cm, clip, keepaspectratio, width=0.4\textwidth]{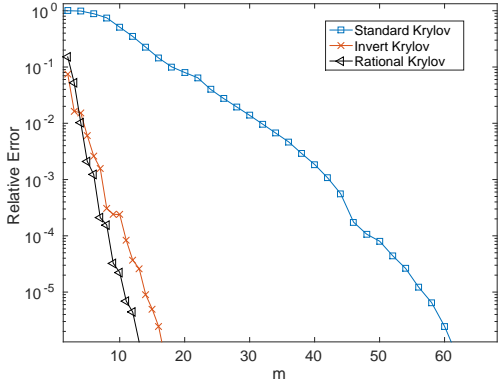}
	\caption{
	{The relative error vs. dimensional $m$ of different Krylov subspace methods. The relative error is $\frac{||e^{h\mbf A}\mbf v - \beta \mbf V_m e^{h\mbf H_m}\mbf e_1||}{||e^{h\mbf A}\mbf v ||}$, where $h=0.4ps$, $\gamma = 10^{-13}$. Note: The relative error is the  difference compared to analytical solution $e^{h\mbf A}\mbf v$ of the ODE $ \frac{d\mbf x}{dt}= \mbf A \mbf x$ with an initial vector $\mbf v$, 
	which is generated by MATLAB $rand$ function, and its entries are positive numbers in $(0,1]$.  			 
	}
}
\label{fig:order_reduction}
\end{figure}

In order to observe the error distribution versus dimensions of standard, invert, and rational Krylov subspace methods for MEVP, we construct a RC circuit with stiffness $$\frac{Re(\lambda_{min})}{Re(\lambda_{max})} = 4.7 \times 10^{6},$$ where $\lambda_{max} = -8.49\times 10^{10}$ and $\lambda_{min} = -3.98 \times 10^{17}$ are the  maximum and minimum eigenvalues of $\mbf A = -\mbf C^{-1} \mbf G$. Fig. \ref{fig:order_reduction} shows the relative error reductions along the increasing Krylov subspace dimension. The error reduction  rate of rational Krylov subspace is the best, while the one of standard Krylov subspace requires huge dimension to capture the same level of error. For example, it costs almost $10\times$ of the size to achieve around relative error $1\%$ compared to Invert and Rational Krylov subspace methods.
The relative error is 
$$\frac{||e^{h\mbf A}\mbf v - \beta \mbf V_m e^{h\mbf H_m}\mbf e_1||}{||e^{h\mbf A}\mbf v ||},$$ 
where $h=0.4ps$, $\gamma = 10^{-13}$. 
The matrix $\mbf A$ is a relatively small matrix and computed by MATLAB $expm$ function. The result of $e^{h\mbf A}\mbf v$ serves as the baseline for accuracy. 
The relative error is the real relative difference compared to the analytical solution $e^{h\mbf A}\mbf v$ of the ODE $$\frac{d\mbf x}{dt}= \mbf A \mbf x$$ with an initial vector $\mbf v$, which is generated by MATLAB $rand$ function. 

The error reduction rate of standard Krylov subspace is the worst, while the  rational Krylov subspace is the best. It is the reason that we prefer rational Krylov subspace (R-MATEX). The relative errors of BE, TR and FE are $0.0594$, $0.4628$, and $2.0701 \times 10^4$, respectively. The large error of FE is due to the instability issue of its low order explicit time integration scheme.
In Fig. \ref{fig:order_reduction}, when $m=3$, standard, invert and rational Krylov subspace methods have $0.8465$, $0.0175$, and $0.0065$, respectively. 
It illustrates the power of matrix exponential method. Our proposed methods are all stable  and can achieve improved error numbers.

\begin{figure}[t]
\centering
\includegraphics[trim = 0.7cm 0.0cm 0.0cm 0.0cm, clip, keepaspectratio, width=0.4\textwidth]{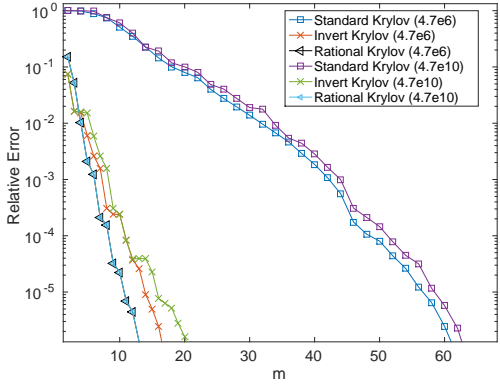}
\caption{{{The relative error vs. dimension $m$ of different Krylov subspace methods. The relative error is $\frac{||e^{h\mbf A}\mbf v - \beta \mbf V_m e^{h\mbf H_m}\mbf e_1||}{||e^{h\mbf A}\mbf v||}$, where $h=0.4ps$, $\gamma = 10^{-13}$. The rational Krylov subspace has very stable error reduction rate. The number in the bracket represents the stiffness value of the system.}}}
\label{fig:order_reduction_2system}
\end{figure}

In order to observe the different stiffness effects on Krylov subspace methods, we change the entries in $\mbf C$ and $\mbf G$ to make the different stiffness value $4.7 \times 10^{10}$. Fig. \ref{fig:order_reduction_2system} illustrates the stable reduction rate of rational method. The stiffness degrades the performance of standard Krylov subspace method.  Both invert and rational Krylov subspace methods are good candidates for stiff circuit system.
	
\begin{figure}[t]
\centering
\includegraphics[trim = 0.1cm 0.0cm 0.0cm 0cm, clip, keepaspectratio, width=0.45\textwidth]{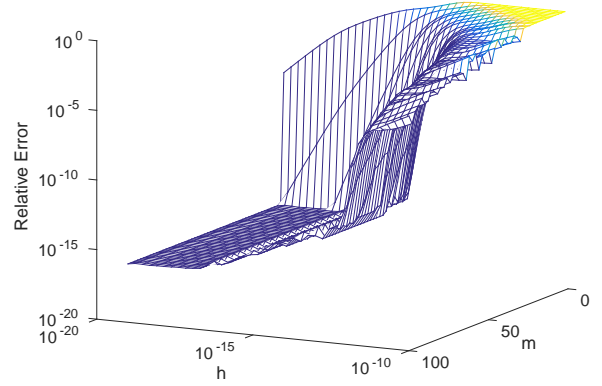}
\caption{The error  of MEVP via standard Krylov Subspace: 
$\frac{||e^{h\mbf A}\mbf v -  \beta \mbf {  V_m} e^{h\mbf {  H}_m}||}{||e^{h\mbf A}\mbf v ||}$ vs. time step $h$ and dimension of standard Krylov subspace basis ($m$). The standard Krylov subspace approximates the solution well in extremely small $h$, since it captures the important eigenvalues and eigenvectors of $\mbf A$ at that region. However, the small $h$ is not useful for the circuit simulation. For large $h$, it costs large $m$ to reduce the error. 
}\label{fig:std}
\end{figure}

\begin{figure}[t]
\centering
\includegraphics[trim = 0.15cm 0.0cm 0.0cm 1cm, clip, keepaspectratio, width=0.45\textwidth]{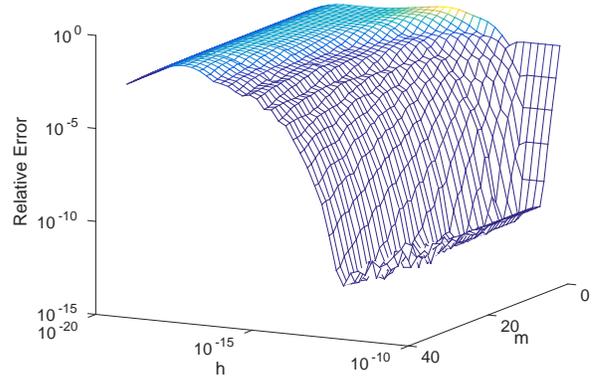}
\caption{
{The error  of MEVP via invert Krylov Subspace:  $\frac{||e^{h\mbf A}\mbf v -  \beta \mbf {V}_m e^{h\mbf {H }^{-1}_m}||}{||e^{h\mbf A}\mbf v ||}$ vs. time step $h$ and dimension of invert Krylov subspace basis ($m$). Compared to Fig. \ref{fig:std}, invert Krylov subspace method reduces the errors for large $h$.}
}
\label{fig:inv}
\end{figure}

\begin{figure}[t]
\centering
\includegraphics[trim = 0.15cm 0.0cm 0.0cm  1cm, clip, keepaspectratio, width=0.45\textwidth]{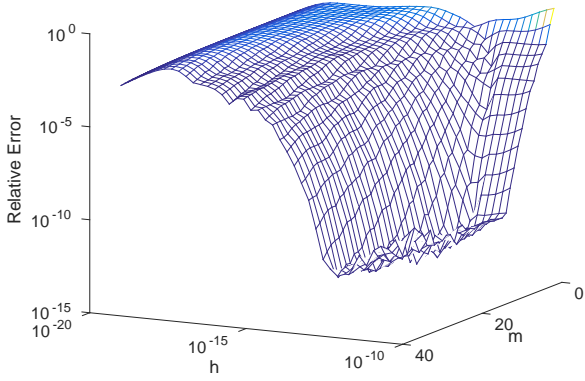}
\caption{
{The error  of MEVP via rational Krylov Subspace: $\frac{||e^{h\mbf A}\mbf v - \beta \mbf {  V}_m e^{h \frac{\mbf I - \mbf { {H}}_m^{-1}}{\gamma}}\mbf e_1||}{||e^{h\mbf A}\mbf v||}$, where $\gamma=5\times 10^{-13}$,  vs. time step $h$ and dimension of rational Krylov subspace basis ($m$). Compared to Fig. \ref{fig:std}, rational Krylov subspace method reduces the errors for large $h$ as Fig. \ref{fig:inv}.}
}
\label{fig:rat}
\end{figure}

Regarding the relative error distributions vs. time step $h$ and dimension $m$, 
Fig. \ref{fig:std}, Fig. \ref{fig:inv}, and  Fig. \ref{fig:rat} are computed by standard, invert, and rational Krylov subspaces ($\gamma = 5\times 10^{-13}$), respectively.
Fig. \ref{fig:std} shows that the errors generated by standard Krylov subspace method has  flat region with high error values in time-step range of interests. The small (`unrealistic') time step range has small error values. 
Compared to Fig. \ref{fig:std}, invert (Fig. \ref{fig:inv}) and rational (Fig. \ref{fig:rat}) Krylov subspace  methods reduce errors quickly for large $h$.
The explanation is that a relatively small portion of the eigenvalues and corresponding invariant subspaces determines the final result (vector) when time step $h$ is larger \cite{Van06}, which are efficiently captured by invert and rational Krylov subspace  methods.


\begin{figure}[t]
	\centering
	\includegraphics[trim = 0cm 0cm 0cm 0cm, clip, keepaspectratio,  width=0.45\textwidth]{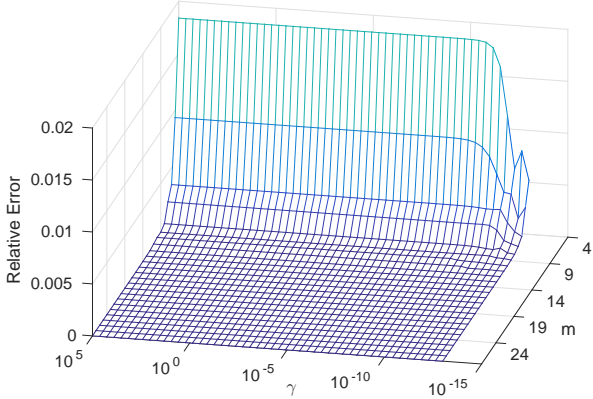}
	\caption{ The error  of MEVP via rational Krylov Subspace 
	{$\frac{||e^{h\mbf A}\mbf v - \beta \mbf {  V}_m e^{h \frac{\mbf I - \mbf { {H}}_m^{-1}}{\gamma}}\mbf e_1||}{||e^{h\mbf A}\mbf v||}$, where $h=4ps$. The flat region shows the error is actually relatively insensitive to $\gamma$, when $\gamma$ is in the range of step size $h$ of interests. 
	}
}
\label{fig:gamma}
\end{figure}

The error of rational Krylov subspace is relatively insensitive to  $\gamma$ when it is selected between the time-step range of interests (Fig. \ref{fig:gamma}).
Above all, rational Krylov (R-MATEX) and invert Krylov (I-MATEX) subspace methods have much better performance than standard version. When we deal with  stiff cases, standard Krylov subspace is not a feasible choice due to the large dimension $m$ of Krylov subspace, which causes huge memory consumption and poor runtime performance. 

\section{MATEX Framework}
\label{sec:matex}
\subsection{MATEX Circuit Solver} 
\label{sec:frame_motivation}
 
We incorporate matrix exponential based integration scheme with Krylov subspace method into our MATEX framework, which is summarized in Algorithm \ref{algo:rmatex}. 
We set $\mbf X_1$ and $\mbf X_2$ in Line \ref{algo:rmatex:init} based on the choice of Krylov subspace method as follows, 
\begin{itemize}
\item I-MATEX: $\mbf X_1 =  \mbf G, ~\mbf X_2=\mbf C$
\item R-MATEX: $\mbf X_1 =  \mbf C+ \gamma \mbf G, ~\mbf X_2=\mbf C$
\end{itemize} 

For linear system of PDN, the matrix factorization in line \ref{algo:rmatex:lu} is only performed once, and the matrices $\mbf L$ and $\mbf U$ are reused in the while loop from line \ref{algo:rmatex:while} to line \ref{algo:rmatex:whileend}. 
Line \ref{algo:rmatex:algo1} uses Arnoldi process  with corresponding inputs to construct Krylov subspace as shown in Algorithm \ref{algo:arnoldi}.

\begin{algorithm}[t]
 	\label{algo:rmatex}
 	{
 		\caption{MATEX Circuit Solver}
 		\KwIn{ $ \mbf C, \mbf G , \mbf B, \mbf u, \epsilon$,
 			and time span $T$. 
 		}
 		\KwOut{  The set of $\mbf x$ from $[0, T]$.}
 		{Set $\mbf X_1, \mbf X_2$\;
 	\label{algo:rmatex:init}
 			$t=0$\;
 			$\mbf x(t) =$DC\_analysis\;  
 			$[\mbf L, \mbf U] = \text{LU\_Decompose}(\mbf X_1)$\; 	\label{algo:rmatex:lu}
 			\While { $t < T$}
 			{\label{algo:rmatex:while}
 				Compute maximum allowed step size $h$\;
 				Update $ \mbf P (t,h),\mbf F (t,h)$\;
 				Obtain  $\mbf x(t+h)$ by  {\bf Algorithm  \ref{algo:arnoldi}} with inputs $[\mbf L, \mbf U, \mbf X_2, h ,t, \mbf x(t), \epsilon, \mbf P (t,h),\mbf F (t,h)]$\; \label{algo:rmatex:algo1}
 				$t=t+h$\;
 			}\label{algo:rmatex:whileend}
 		}
 	}
\end{algorithm}

\subsection{DR-MATEX (Distributed R-MATEX Framework) by Decomposition of Input Sources, Linear Superposition, and Parallel Computation Model}
\label{sec:dist_frame}
\subsubsection{Motivation} \

There are usually many input sources in PDNs as well as their transition activities, which might narrow the regions for the stepping of matrix exponential based method due to the unaligned breakpoints. 
In other words, the region before the next transition $t_s$ may be shortened when there are a lot of activities from the input  sources. 
It leads to more chances of generating new Krylov subspace bases. 
We want to reduce the number of subspace generations and improve the runtime performance.\footnote{The breakpoints also put the same constraint on TR-FTS and  BE-FTS. 
However, their time steps are fixed already, which refrains them from reaching this problem in the first place.}
\vspace{+0.07in}

\subsubsection{Treatment and Methodology} \ 
\label{ltsgts}

In matrix exponential based integration framework, 
we can choose any time spot  $t+h \in [t,t_s]$ with computed Krylov subspace basis. The solution of $\mbf x(t+h)$ is computed by scaling the existing Hessenberg matrix $\mbf H$ with the time step $h$ as below
\begin{eqnarray}
\mbf x(t+h) = \lVert \mbf v \rVert \mbf {V}_m e^{h \mbf H} \mbf e_1-\mbf P(t,h).
\label{sq:snapshot}
\end{eqnarray}
This is an important feature for computing the solutions at intermediate time points without generating the Krylov subspace basis, when there is no current transition. 
Besides, since the PDN is linear dynamical system, we can utilize the well-known superposition property of linear system and distributed computing model to tackle this challenge. 

To illustrate our distributed version of MATEX framework, we first define three terms to categorize the breakpoints of input sources:
\vspace{+0.08in}
\begin{itemize}
	\item 
	\emph{Local Transition Spot} ($LTS$): the set of $TS$ at an input source to the PDN.
	\item
	\emph{{Global Transition Spot}} ($GTS$): the union of $LTS$ among all the input sources to the PDN. 
	\item  
	\emph{{Snapshot}}:  a set $GTS \setminus LTS$ at {one input source}. 
\end{itemize} 
\vspace{+0.08in}

\begin{figure}[t] 
	\centering
	\includegraphics[width=2.3in]{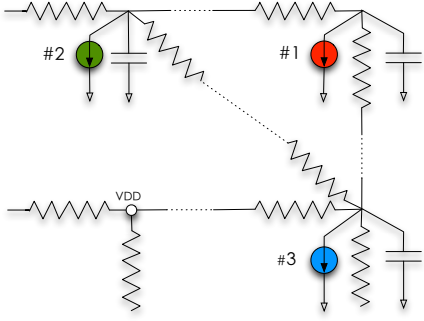}
	\caption{Part of a PDN model with input sources from Fig. \ref{fig:input}.}
	\label{fig:rc_ckt}
\end{figure}

\begin{figure}[ht] 
	\centering
	\includegraphics[width=3.4in]
	{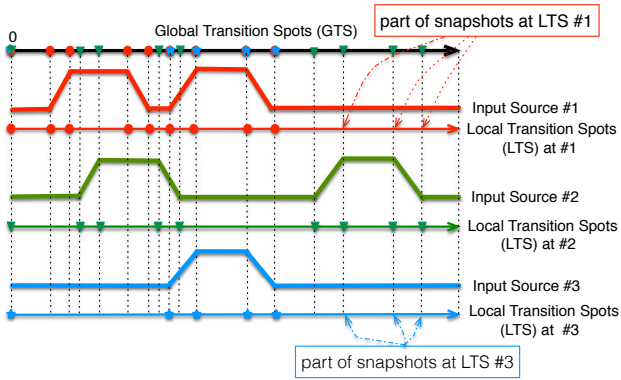} 
	\caption{ 
		{Illustration of input transitions. 
			$GTS$: Global Transition Spots; $LTS$: Local Transition Spots;
			{$Snapshots$: the crossing positions by dash lines and LTS \#$k$ without solid points.} 
		}
	}
	\label{fig:input}
\end{figure}

If we simulate the PDN with respect to all the input sources, the points in the set of $GTS$ are the places where generations of Krylov subspace cannot be avoided. For example, there are three input sources in a PDN (Fig. \ref{fig:rc_ckt}). The input waveforms are shown in Fig. \ref{fig:input}. The first line is $GTS$, which is contributed by the union of $LTS$ in input sources \#1, \#2 and \#3. However, we can partition the task into subtasks by simulating each input source individually. Then, each subtask generates Krylov subspaces based on its own $LTS$ and keeps track of $Snapshot$ for the later usage of summation via linear superposition. 
Between two LTS points $t$ and $t+h$,  the \emph{Snapshot} points $$t+h_1 < t+h_2 < \cdots < t+h_l  \in (t, t+h]$$ can reuse the Krylov subspace generated at $t$. 
For each node, the chances of generation of Krylov subspaces are reduced. 
The time periods of reusing latest Krylov subspaces are enlarged locally and bring the runtime improvement.
Besides, when subtasks are assigned, there is no communication among the computing nodes, which leads to so-called \emph{Embarrassingly Parallel} computation model.  
\vspace{+0.07in}

\begin{figure}[t] 
	\centering
	\includegraphics[width=3.4in]
	{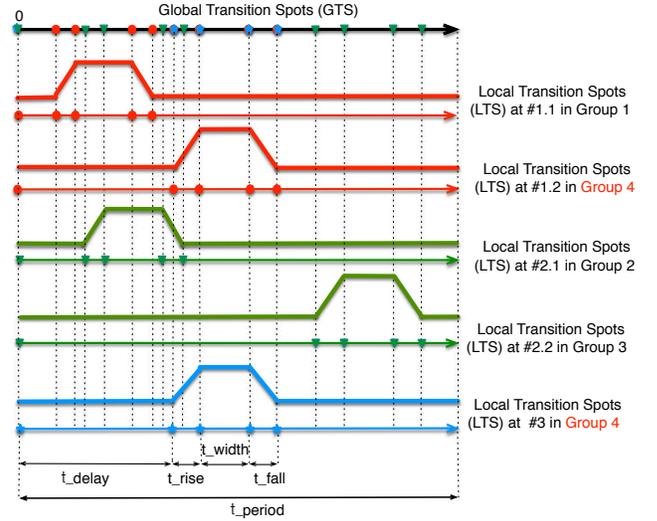}
	\caption{Grouping of ``Bump'' shape transitions for sub-task simulation.
		The matrix exponential based method can  utilize  adaptive stepping in each $LTS$ and reuse the Krylov subspace basis generated at the latest point in $LTS$. 
		However, traditional methods (TR, BE, etc.) still need to do time marching, either by pairs of  forward and backward substitutions and proceed with fixed time step, or by re-factorizing matrix and solving linear system for adaptive stepping.
		(Pulse input information:
		$t_{delay}$: delay time;
		$t_{rise}$: rise time;
		$t_{width}$: pulse width; 
		$t_{fall}$: fall time; and
		$t_{period}$: period).
	}
	\label{fig:ckt_ts_decomp}
\end{figure}

\begin{figure*}[t]
	\centering
	\includegraphics[width=0.6\textwidth]
	{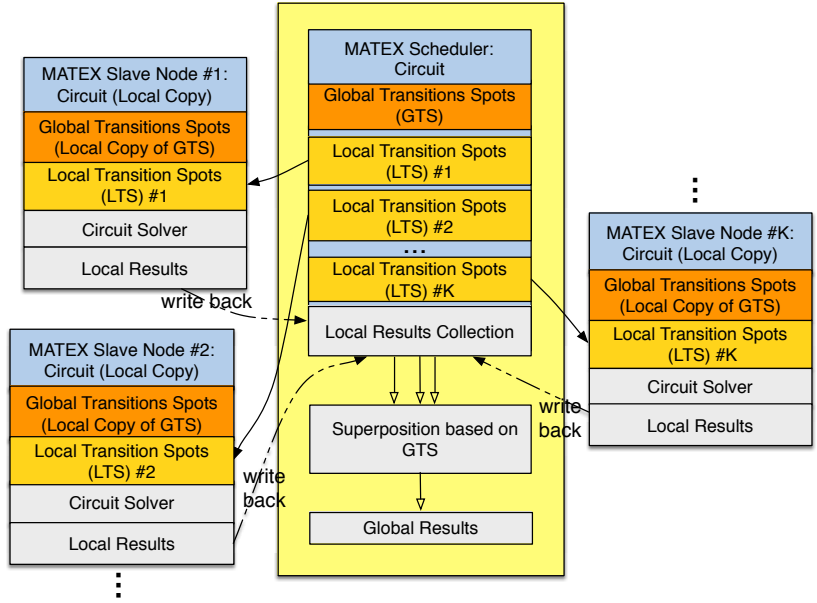}
	\caption{DR-MATEX: The distributed MATEX framework using R-MATEX circuit solver.}
	\label{fig:drmatex}
\end{figure*} 

\subsubsection{More Aggressive Tasks Decomposition} \ 
\label{bump}
We divide the simulation task based on the alignments of input sources. More aggressively, we can  decompose the task according to the ``bump'' shapes of the input sources.\footnote{IBM power grid benchmarks provide the pulse input model in SPICE format.}  We group the input sources, which have the same $$(t_{delay}, t_{rise}, t_{fall}, t_{width})$$ into one set. 
For example, the input source \#1 of Fig. \ref{fig:input} is divided to  \#1.1 and \#1.2 in Fig. \ref{fig:ckt_ts_decomp}. 
The input source \#2 in Fig. \ref{fig:input} is divided to  \#2.1 and \#2.2 in Fig. \ref{fig:ckt_ts_decomp}.  Therefore, there are four groups in Fig. \ref{fig:ckt_ts_decomp}, Group 1 contains $LTS\#1.1$. Group 2 contains $LTS\#2.1$. Group 3 contains $LTS \#2.2$. 
Group 4 contains $LTS \#1.2$ and $\#3$. Our proposed framework MATEX is shown in Fig. \ref{fig:drmatex}. After pre-computing $GTS$ and decomposing $LTS$ based on ``bump'' shape (Fig. \ref{fig:ckt_ts_decomp}), we group them and form $LTS$ $\#1 \sim \#K$.\footnote{There are alternative decomposition strategies. It is also easy to extend the work  to deal with different input waveforms. We try to keep this part as simple as possible to emphasize our framework.} 
\vspace{+0.07in}

\subsubsection{MATEX Scheduler in DR-MATEX} \

In DR-MATEX, the role of MATEX scheduler is just to send out $GTS$ and $LTS$ to different MATEX slave nodes and collect final results after all the subtasks of transient simulation are finished. 
The node number is based on the total number of subtasks, which is the group number after PDN source decomposition. 
Then the simulation computations are performed in parallel. 
Each node has its own  inputs. For example, Node\#$k$ has $GTS$, $LTS$\#$k$, $\mbf P_k$ and $\mbf F_k$, which contain the corresponding $\mbf b$ for node $k$.
Scheduler does not need to do anything during the transient simulation, since there are  no communications among nodes before the stage of ``write back'' (in Fig. \ref{fig:drmatex}), by when all nodes complete their transient simulations.

Within each slave node, the circuit solver (Algorithm \ref{algo:ckt_solver}) computes transient response with varied time steps. Solutions are obtained without re-factorizing matrix during the computation of transient simulation. The computing nodes write back the results and inform the MATEX scheduler after finishing   their own transient simulation.

\begin{algorithm}
	\label{algo:ckt_solver} 
	{ 
		\caption{DR-MATEX: The distributed MATEX framework using R-MATEX at Node\#$k$.} 
		\KwIn{ $LTS$\#$k$, $GTS$, $\mbf P_k$, $\mbf F_k$, error tolerance $E_{tol}$, and simulation time span $T$.}
		\KwOut{Local solution $\mbf x$ along $GTS$ in node 
			$k \in [1,\cdots,S]$, where $S$ is the number of nodes}
		{   
			$t=0$, $\mbf X_1= \mbf C + \gamma \mbf G$, and $\mbf X_2 = \mbf C$\;
			$\mbf x(t) = \text{Local\_Initial\_Solution}$\;
			$[\mbf L,\mbf U] = \text{LU\_Decompose}( \mbf X_1)$\;
			\While { $t < T$}
			{
				Compute maximum allowed step size $h$ based on $GTS$\;
				{
					\If{$t \in $  LTS\#$k$}
					{
						\tcc{\small  Generate Krylov subspace for the point at $LTS\#k$ and  compute $\mbf x(t+h)$}
						$[\mbf x(t+h), \mbf V_m, \mbf H_m,\mbf v] = \text{MATEX\_Arnoldi}(\mbf L,\mbf U,\mbf X_2,$ $ h, t, \mbf x(t), \epsilon, \mbf P_k(t,h), \mbf F_k(t,h))$\;
						$a_{lts} = t$\; 
					}
					\Else{
						\tcc{\small Obtain $\mbf x(t+h)$ at $Snapshot$ with computed Krylov subspace}
						{
							$h_{a} = t + h - a_{lts}$\;
							$\mbf x(t+h) = \lVert \mbf v \rVert \mbf {V}_m e^{h_{a} \mbf H_m} \mbf e_1 - \mbf P_k(t,h)$\;
	     				}
					}
					$t= t+h$\;
				}
			}    }
		}
	\end{algorithm}
	
\subsection{Runtime Analysis of MATEX PDN Solver } 
\label{sec:complexity} 

Suppose we have the dimension of Krylov subspace basis $m$ on average for each time step and one pair of forward and backward substitutions consumes runtime $T_{bs}$. 
The total time of serial parts is $T_{serial}$, which includes matrix factorizations, result collection, etc. 
For $\mbf x(t+h)$, the evaluation of matrix exponential with $e^{h\mbf H_m}$ is $T_{H}$, which is in proportion to the time complexity $O(m^3)$. Besides, we need extra $T_e$ to form $  \mbf x(t+h)$, which is proportional to $O(n m^2)$ by $\beta\mbf V_m e^{h\mbf H_m} \mbf e_1$. 

Given $K$ points of $GTS$, without decomposition of input sources, the runtime is \begin{eqnarray} 
\label{eq:full}
KmT_{bs} + K (T_{H}+ T_e)  + T_{serial}.
\end{eqnarray}
After dividing the input transitions and sending to enough computing nodes, we have $k$ points of $LTS$ for each node based on feature extraction and grouping (e.g., $k=4$ for one ``bump'' shape feature). The total computation runtime is  
\begin{eqnarray} 
\label{eq:reduced}
kmT_{bs} + K (T_{H}+ T_e)  + T_{serial},
\end{eqnarray}
where  $K (T_{H}+ T_e)$ contains the portion of computing $Snapshot$ in DR-MATEX mode. 
The speedup of DR-MATEX over single MATEX is  
\begin{eqnarray}
\text{Speedup}=\frac{ KmT_{bs} + K(T_{H}+ T_e)+ T_{serial}} { kmT_{bs}+ K(T_{H}+ T_e)+T_{serial}}.
\end{eqnarray}

For R-MATEX, we have  small $m$. Besides, $T_{bs}$ is relatively larger than $(T_{H}+T_e)$ in our targeted problem. 
Therefore, the most dominating part is the $KmT_{bs}$ in Eq. (\ref{eq:full}). We can always decompose input source transitions, and make $k$  smaller than $K$. 

In contrast, suppose the traditional method with fixed step size has $N$ steps for the  entire simulation, the runtime is $$NT_{bs}+T_{serial}.$$ 
Then, the speedup of distributed DR-MATEX over the traditional method is 
\begin{eqnarray}
	\label{eqn:tr_spd}
	\text{Speedup}'=\frac{NT_{bs}+T_{serial} }  { kmT_{bs}+ K(T_{H}+T_e)+T_{serial}}.
\end{eqnarray}

Note that, when the minimum distance among input source breakpoints decreases, large time span or many cycles is required to simulate PDNs, the schemes with such uniform step size would degrade runtime performance furthermore due to the increase of $N$. 
In contrast, in MATEX PDN solver, $K$ is not so sensitive to such constraints. Besides, $k$ can be maintained in a small number based on the decomposition strategy. 
Therefore, the speedups of our proposed methods tend to be  larger when the simulation requirements become harsher. 

 
\section{Experimental Results}
\label{sec:exp_results} 
We implement all the algorithms 
in MATLAB R2014b\footnote{Measurements reported are on MATLAB implementations. They are subject to limitations and are not directly comparable to C++ implementations reported in literature such as \cite{Wang13}.} and use UMFPACK package for LU factorization. 
First, we compare I-MATEX, R-MATEX and TR in order to show our runtime improvements in single machine framework in Table \ref{tab:perf_mexp}. 
Second, we show our distributed framework DR-MATEX achieves large speedups in Table \ref{tab:pg_rmatex}. 
The experiments are conducted on the server with Intel(R) Xeon (R)~E5-2640 v3 2.60GHz processor and 125GB memory.

\subsection{Performance of I-MATEX and R-MATEX in Sec. \ref{sec:frame_motivation}}
\label{sec:resultI}
We compare our proposed I-MATEX and R-MATEX against the popular TR-FTS on the IBM power grid benchmarks \cite{Nassif08_Power}. Among the current sources, the smallest interval between two breakpoints is $h_{upper}=10ps$, which puts the upper limit of the TR's step size. All of these cases have very large numbers of input current sources. Table~\ref{tab:spec_ibmpg} shows the details of each benchmark circuit of which size ranges from $54$K up to $3.2$M. The simulation time is $10ns$. From ibmpg1t to ibmpg6t, TR uses fixed step size in $10ps$. We also change the IBM power grid benchmark to make the   smallest distance among breakpoints  $1ps$ by interleaving input sources' breakpoints (similar as Fig. \ref{fig:interleave}). Therefore, the fixed step size method can only use  at most $1ps$. The names of those benchmarks are ibmpg1t\_new, ibmpg2t\_new, ibmpg3t\_new, ibmpg4t\_new, ibmpg5t\_new and ibmpg6t\_new.  

\begin{table}[t]
\centering
\caption{Specifications of IBM power grid benchmarks.}
\begin{tabular}{|c|r|r|r|r|r|r|r|} 
\hline
Design    & \#R     & \#C      & \#L  & \#I    & \#V  & \#Nodes  \\ \hline
ibmpg1t   & 41K   & 11K    & 277  & 11K  & 14K  & 54K  \\ \hline
ibmpg2t   & 245K  & 37K    & 330  & 37K  & 330    & 165K  \\  \hline
ibmpg3t   & 1.6M & 201K   & 955  & 201K  & 955    & 1.0M \\ \hline
ibmpg4t   & 1.8M & 266K   & 962  & 266K  & 962    & 1.2M \\ \hline
ibmpg5t   & 1.6M & 473K   & 277  & 473K  & 539K   & 2.1M \\ \hline
ibmpg6t   & 2.4M & 761K   & 381  & 761K &836K  & 3.2M \\ \hline
\end{tabular}
\label{tab:spec_ibmpg}
\end{table} 

After DC analysis in TR-FTS, we LU factorize matrix once for the  later transient simulation, which only contains time stepping. 
Actually, multiple factorized matrices can be deployed \cite{Li11, Ye08_ISQED}. 
We can choose one of them during the stepping. The problem is the memory and runtime overhead for the multiple matrix factorizations. 
Another point is if large time step $h'$ is chosen, the standard low order scheme cannot maintain the accuracy. 

Experiment is conducted  on a single computing node.  In Table \ref{tab:perf_mexp}, we record the total simulation runtime {\bf{Total(s)}}, which includes the processes of DC and transient simulation, but excludes the non-numerical computation before DC, e.g., netlist parsing and matrix stamping.  
We also record the part of transient simulation {\bf{Tran(s)}}, excluding DC analysis and LU decompositions. 
The speedup of I-MATEX is not as large as R-MATEX, because I-MATEX with a large spectrum of $\mbf A$ generates large dimension $m$ of Krylov subspace.  
Meanwhile, the step size is not large enough to let it fully harvest the gain from time marching with stepping.
In contrast, R-MATEX  needs small dimension numbers $m$ of rational Krylov subspace, which  ranges from $2$ to $8$ in those cases. Therefore, they can benefit from large time stepping, shown as SPDP$^{r}_{tr}$. For ibmpg4t, R-MATEX achieves maximum speedup resulted from the relatively small number of breakpoints in that benchmark, which is around $44$ points, while the majority of  others  have over $140$ points.  

In Table \ref{tab:perf_mexp}, our single mode R-MATEX achieves the average speedup {\bf $5\times$} over TR-FTS. Note the average speedup number of single mode R-MATEX over TR-FTS for the original IBM benchmark (ibmpg1t$\sim$ibmpg6t) is less than the speedup of the new test cases (ibmpg1t\_new$\sim$ibmpg6t\_new). As we mentioned before, ibmpg1t\_new$\sim$ibmpg6t\_new have harsher input constraints, making the available step size only $1ps$.  Therefore, the adaptive stepping by R-MATEX is more beneficial to the runtime performance in ibmpg1t\_new$\sim$ibmpg6t\_new than ibmpg1t$\sim$ibmpg6t. 

\begin{table*}[t]
	\centering
	 	\caption{
		 {Performance comparisons (single computing node): TR-FTS, I-MATEX, and R-MATEX. 
		 {\bf DC(s)}: Runtime of DC analysis (seconds);
		 {{\bf $m_I$}: The maximum $m$ of Krylov subspace in I-MATEX.}
		 {\bf Tran(s)}: Runtime of transient simulation after DC (seconds), excluding the matrix factorization runtime; 
		 {\bf Total(s)}: Runtime of overall transient simulation (seconds); \
		  {{\bf Df(uV)}: Maximum and average voltage differences compared to provided solutions (uV); } 
		 { {\bf  $m_R$}: The maximum $m$ of Krylov subspace in R-MATEX}
		 {\bf SPDP$^{r}_{tr}$}: Speedup of R-MATEX over TR-FTS with respect to {\bf Tran(s)};
		  {\bf SPDP$^{r}_{i}$}:  Speedup of R-MATEX over I-MATEX with respect to {\bf Tran(s)}. 
	}
	}
	\begin{threeparttable}[b]
		\begin{tabular}{|c|r|r|r||r|r|r|r||r|r|r| r||r|r| 
				}
			\hline
			\multirow{2}{*}{Design} & 
			\multirow{2}{*}{DC(s)} &
			\multicolumn{2}{c||}{TR-FTS} & 
			\multicolumn{4}{c||}{I-MATEX} & 
			\multicolumn{4}{c||}{R-MATEX} & 
			\multicolumn{2}{c|}{Speedups} 
			\\ 
			\cline{3-14} 
			&	
			& Tran(s)	&Total(s)	
			& $m_I$	
			& Tran(s)
			& Total(s) 
			& Df(uV)
			& $m_R$	
			& Tran(s)
			& Total(s) 
			& Df(uV)
			& SPDP$^{r}_{tr}$ 
			& SPDP$^{r}_{i}$ 
			\\ \hline  \hline 
			ibmpg1t		     
			& 0.2  & 5.7 	& 6.00 
			& 30  	& 28.8 	& 28.9	& $58 \backslash 9.8$  
			& 5 	& 10.1 & 10.3  & $45 \backslash 6.8 $ & $0.6\times$ 
			& $2.9\times$
			\\ \hline 
			ibmpg2t    
			& 0.8  & 40.0 & 41.9 	
			& 28 & 130.0 &  130.9	& $92  \backslash 10.5$
			& 5  & 35.6 & 37.4 & $45 \backslash 6.8$ 
			&  $1.1\times$	&  $ 3.7\times$ 
			\\ \hline 
			ibmpg3t 	     
			&  16.4  &  263.2  	& 295.0  
			& 29 & 1102.5  & 1115.1 	& $95 \backslash 20.4$ 
			&  5 & 	275.5  & 301.0 	&  $95 \backslash 18.5$
			& 	$1.0\times$ 	&	$4.0 \times$		
			\\ \hline
			ibmpg4t    
			& 13.5 & 460.8 	& 501.9
			&  29 
			 & 433.8	& 458.2 &  $101  \backslash 39.3$
			&  5& 200.5  & 239.1   & $99 \backslash 34.2$ 
			 & $ 2.3\times$
			& 	 $ 2.2\times$ 
			\\ \hline
			ibmpg5t   
			&  9.0 	&  476.6 	&  	498.0 
			&  30 &  1934.4	& 1944.5  	&  $29  \backslash 5.6$		
			&  5 & 383.1   & 401.9 &  $29  \backslash 4.4$    
			&  $ 1.2\times$
			& $ 5.0\times$	
			\\ \hline 
			ibmpg6t    
			&  15.3 	&  716.0 	& 749.1  	
			& 25  &  2698.9	& 2713.7  	&   $39\backslash 8.6$
			& 5 & 773.5 & 800.5 	&   $33\backslash 5.6$	
	&  $0.9\times$ 
			& $3.5\times$ 
			\\ 
			\hline
			ibmpg1t\_new	   
			&  0.2	&  51.3 & 51.7
			& 30 &  27.2  & 27.4 &  $58  \backslash 9.8$	  
			& 5 & 11.7   & 12.1  & $53 \backslash 6.9 $	 
			& $4.4\times$ 	& $2.3\times$
			\\ \hline 
			ibmpg2t\_new 	      
			&  0.9   &  431.4 	&  	433.5
			&  28 	& 114.9  & 115.7 	&  $49 \backslash 10.5$	   
			& 5 	& 43.3 	& 44.9  &  $33\backslash 5.6$	  
			& $10.0\times$ 		& $2.7\times$	 
			\\ \hline 
			ibmpg3t\_new	        
			&  16.3 &  3716.5  	&  	3749.0
			&  29	&  1219.3 & 1232.6  	&  $95 \backslash 20.4$	
			&  5	& 481.7   & 508.2 & $95 \backslash 18.9$	    
			&  $7.7\times$ 
			&  $2.5\times$	 
			\\ \hline
			ibmpg4t\_new 	     
		 	& 18.3 	& 5044.6 & 5085.3 
		 	&  29   
		 	& 753.5 	& 776.4 	&  $101 \backslash 39.3$		 
			& 6	& 350.9 & 387.2 & $99 \backslash 34.2$		
			&  ${14.4}\times$ & 	 $2.1 \times$ 
			\\ \hline
			ibmpg5t\_new 	     
			& 10.5  & 5065.9 & 5110.1
			& 30  	& 2494.0	&  2504.7
			& $ 30 \backslash 5.6 $ 
			& 5  & 746.2	&  766.4 & $ 30 \backslash 4.4 $   &  $6.8\times$			
			&  $3.3\times$	  
			\\ \hline 
			ibmpg6t\_new 	       
			&  13.1	&  7015.3 	&  	7059.7
			& 25 & 3647.9 & 3663.1 &  $39 \backslash 8.6 $ 
			& 6	& 895.1	&  923.1 &  $33\backslash 7.3 $ 
            &  $7.8\times$  &	$4.1\times$	 
			\\ 
			\hline	
			\hline
			 Average   & --- & ---
			  & --- & --- & --- & --- 
			  & $65\backslash 15.7$ &  --- & ---
			 &  --- & $57\backslash 12.8$ 
		 & $5\times$
			 &$3\times$ 
			 \\
			 \hline 		
		\end{tabular} 
	\end{threeparttable}
	\label{tab:perf_mexp}
\end{table*}

\subsection{Performance of DR-MATEX in Sec. \ref{sec:dist_frame}}  
 
We test our distributed DR-MATEX in the following experiments with the same IBM power grid benchmarks. These cases have many input transitions ($GTS$) that limit step sizes of R-MATEX. 
We divide the region before the computation of simulation. 
We decompose the input sources by the approach discussed in Sec. \ref{bump} and obtain much fewer transitions of $LTS$ for computing nodes.
The  original input source  numbers are over ten thousand in the benchmarks. 
However, based on ``bump'' feature (as shown in Fig. \ref{fig:ckt_ts_decomp}), we obtain a fairly small numbers for each computing node, which is shown as \emph{Group \#} in Table  \ref{tab:pg_rmatex}. 
(Now, the fact that hundred machines to process in parallel is quite normal \cite{mesos, cloud} in the industry.)
We pre-compute $GTS$ and $LTS$ groups and assign sub-tasks to corresponding nodes\footnote{ 
Based on the feature of input sources available, the preprocessing is very efficient, which takes linear time complexity to obtain GTS, LTS and separates the  sources into different groups.}.
MATEX scheduler is only responsible for simple superposition calculation at the end of simulation.
Since the slave nodes are in charge of all the computing procedures (Fig. \ref{fig:drmatex}) for the computation of their own transient simulation tasks, and have no communications with  others, our framework falls into the category of \emph{Embarrassingly Parallelism} model. We can easily emulate the multiple-node environment.  
We simulate each group using the command ``matlab -singleCompThread'' in our server. We record the runtime numbers for each process (slave nodes) and report the maximum runtime as the total runtime ``Total(s)'' of DR-MATEX in Table \ref{tab:pg_rmatex}. We also record ``pure transient simulation'' as ``Tran(s)'', which is the maximum runtime of the counterparts among all computing nodes. 

For TR-FTS, we use $h=10ps$, so there are 1,000 pairs of forward and  backward substitutions during the process of pure transient simulation for ibmpg1t$\sim$ibmpg6t; 
We use $h=1ps$ for ibmpg1t\_new$\sim$ibmpg6t\_new. 
Therefore, we have 10,000 pairs of forward and  backward substitutions for stepping.
In DR-MATEX, the circuit solver uses R-MATEX with {\bf $\gamma = 10^{-10}$}, which is set to sit among the order of varied time steps during the simulation (since Sec. \ref{sec:compare_krylov} discusses the insensitivity of $\gamma$ around the step size of interests). 
TR-FTS is not distributed because it has no gain by dividing the current source as we do for the DR-MATEX. TR-FTS cannot avoid the repeated pairs of forward and backward substitutions. Besides, adaptive stepping for TR-FTS only degrades the performance,
since the process requires extra matrix factorizations.

In Table \ref{tab:pg_rmatex}, our distributed mode gains {{up to  $98\times$}}   for the pure transient computing. The average peak  dimension $m$ of rational Krylov subspace is $7$. The memory overhead ratio for each node (around $1.6 \times$  over TR-FTS on average) is slightly larger, which is worthwhile  with respect to the large runtime improvement.  
With the huge reduction of runtime for Krylov subspace generations, the serial parts, including LU and DC, play more dominant roles in DR-MATEX, which can be further improved using advance matrix solvers, such as \cite{nicslu}.

\begin{table*}[t] 
	\caption{
		{ 
			{The performance of DR-MATEX (Distributed R-MATEX)}. {\bf Group \#}: Group number of the testcases. 
			This number represents the total number of simulation sub-tasks for the design;
			{\bf Tran(s)}: Runtime of transient simulation after DC (seconds); 
			{\bf Total(s)}: Runtime of overall transient simulation (seconds); 
			{\bf Max. Df.(V)}  and {\bf Avg. Df.(V)}: maximum and average differences compared to the solutions of all output nodes provided by IBM power grid benchmarks.  
			{\bf SPDP$_{tr}$}: Speedup over TR-FTS's {\bf Tran(s)} in Table \ref{tab:perf_mexp};
			{\bf SPDP$_r$}: Speedup over R-MATEX's {\bf Tran(s)} in Table \ref{tab:perf_mexp};
			{\bf Peak $m$}: the peak dimension used in DR-MATEX  for MEVP; 
			{\bf Mem. Ratio over TR-FTS}: The peak memory comparison between the maximum memory consumption of DR-MATEX  over TR-FTS in Table \ref{tab:perf_mexp}.
		}
	} 
	\centering
	\begin{tabular}{|c||r|r|r|r|r||r|r||c|c|}
		\hline
		\multirow{2}{*}{Design}&
		\multicolumn{5}{c||}{DR-MATEX}&
		\multicolumn{2}{c||}{Speedups}
		& Peak   &
		{Mem. Ratio} 
		\\
		\cline{2-8}
		& Group \#
		& Tran(s)
		& Total(s)
		& Max Df.(V)
		& Avg Df.(V)
		& SPDP$_{tr}$
		& SPDP$_{r}$
		& $m$
		& over TR-FTS
		\\
		\hline \hline 
		ibmpg1t & 100  &1.4  &  1.9 
		&  5.3e-5   & 8.6e-6   
		& $4.0\times$ 
		&  $7.1\times$      & 6 & 1.9
		\\
		\hline
		ibmpg2t & 100 & 8.9
		& 11.4 & 
		4.6e-5   & 8.6e-6    
		&  $4.5\times$  
		&   $4.0\times$  
		& 7 & 1.9 
		\\
		\hline
		ibmpg3t & 100 & 91.7  & 129.9
		&  9.6e-5 & 19.7e-6   &  $2.9\times$ 
		&   $4.4\times$  & 6  &  1.5 
		\\
		\hline
		ibmpg4t & 15  & 52.3 & 112.2	&  9.9e-5  & 27.9e-6    &  $8.8\times$   &    $3.8\times$  &  8 & 1.4  
		\\
		\hline   
		ibmpg5t & 100  	& 148.4 & 178.9  		&  9.0e-5  &  1.1e-6     
		&    $3.2\times$ 
		&   $2.6\times$   &  7 
		& 1.5
		\\
		\hline
		ibmpg6t  & 100    
	& 189.9  & 234.2	&  3.4e-5  &  7.2e-6   
		&  $3.8\times$    
		& $4.1\times$   & 7 
		& 1.5 
		\\
		\hline  
		ibmpg1t\_new  &  {\color{black} 100} 
		&  2.4   & 2.8 
		&  5.3e-5   & 8.6e-6     
		&  $21.8\times$  
		& $5.0\times$  & 6  & 1.9
		\\
		\hline
		ibmpg2t\_new &{\color{black} 100}  
		& 5.6 & 7.0
		& 4.6e-5 & 8.6e-6  
		& $61.6\times$   
		 & $6.2\times$ &  7 & 1.9
		\\
		\hline
		ibmpg3t\_new  &  {\color{black} 100} & 	103.0 & 140.9 
		&  9.8e-5 & 19.9e-6 
		&  $25.6\times$  
		 &  $3.3\times$  
		& 7  & 1.5 
		\\
		\hline
		ibmpg4t\_new & {\color{black} 15}   
	& 51.5 & 108.4
		&  9.9e-5  & 27.6e-6 
		& {$98.0\times$}  
		  &  $6.8\times$  & 8   & 1.4 
		\\
		\hline
		ibmpg5t\_new  &  {\color{black} 100}  
		& 185.6 & 227.8
	 & 9.9e-5  &  2.2e-6     
		& $27.3\times$ 
		 & $4.0\times$ 
		& 7 &  1.5
		\\
		\hline
		ibmpg6t\_new  & {\color{black} 100}   
	&  274.8  
&  317.7
		&  3.4e-5  &  7.1e-6
         &  $25.5\times$  
		 &  $3.3\times$   &  7  & 1.5   
		\\
		\hline 	\hline
		 {Average}
		& --- & --- & --- &  {7.1e-5} &  {12.3e-6}   
		& { $26\times$}
		 &  { $5 \times$} &  {6.7} 
		& 
		 {1.6}
		\\
		\hline 
	\end{tabular}
	\label{tab:pg_rmatex}
\end{table*}

\section{Conclusions and Future Directions} 
\label{sec:conclusion}

In this work, we propose an efficient framework MATEX for accurate PDN time-domain simulation based on the  exponential integration scheme. We visualize the error distributions to show the advantages of using rational (R-MATEX) and invert (I-MATEX) Krylov subspace methods for matrix exponential and vector product (MEVP) over standard Krylov subspace method (MEXP). For the PDN simulation, our time integration scheme can perform adaptive time stepping without repeating matrix factorizations, which cannot be achieved by traditional methods using implicit numerical integration with fixed time-step scheme. Compared to the commonly adopted framework TR with fixed time step (TR-FTS), our single mode framework (R-MATEX) gains runtime speedup {\color{black} up to around $15\times$}. We also show that the distributed MATEX  framework (DR-MATEX) leverages the superposition property of linear system and  decomposes the task based on the feature of input sources, so that we  reduce chances of Krylov subspace generations for each node. We achieve runtime improvement {up to   $98\times$} speedup.

We show the exponential integration with Krylov subspace methods maintains high order accuracy and flexible time stepping ability. 
The exponential integration framework was actually mentioned by the very early work in circuit simulation algorithms \cite{Leon75}, but it had not attracted too much attention due to the high computational complexities of matrix exponential during that time.  Nowadays,  the progress of Krylov subspace methods provides efficient way to compute matrix exponential and vector product, so that we can utilize certain features of exponential integration, which are hardly obtained by traditional time integration schemes. 
Exponential integration can also serve as stable explicit schemes \cite{Hochbruck10_EXP, Hochbruck98_JSC} for general dynamical systems. 
It is a promising framework for the future  circuit simulation algorithms and software. 
The opportunities of parallel and distributed computing with the cutting-edge multi-core and many-core hardware are also worth exploring for the further parallelism and runtime improvement.


\section*{Acknowledgment}
We thank Prof. Scott B. Baden,  Prof. Mike Botchev, Dr. Quan Chen, Dr. Peng Du,  Dr. Martin Gander, Prof. Nicholas Higham, Prof. Marlis Hochbruck,  Junkai Jiang,  Dr. John Loffeld, Dr. Jingwei Lu,  Mulong Luo, Prof. Alexander Ostermann, Prof. Mayya Tokman    and  Chicheng Zhang, for the informative discussions.  
Hao Zhuang thanks the supports from UCSD's Powell Fellowship and Qualcomm FMA Fellowship.
Last but not the least, we thank for all the insightful comments from the reviewers.
  
\bibliographystyle{ieeetr}
\bibliography{CADJournal_ljw} 
\begin{IEEEbiography}[{\includegraphics[width=1in,clip,keepaspectratio]{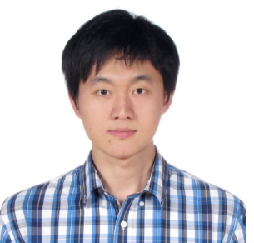}}]{Hao Zhuang}(S'11) is a Ph.D. candidate at the Department of Computer Science and Engineering, University of California, San Diego, CA, USA (UC San Diego). He received his C.Phil. degree in Computer Science from UC San Diego in June, 2015. His current research interests include numerical and optimization algorithms, with the applications in design automation, very large-scale integration systems, network and data analysis.
	 
He has industrial experiences in several companies, including Synopsys, Inc., Mountain View, CA, USA, and Ansys, Inc., San Jose, CA, USA, where he worked on the large-scale circuit analysis and  dynamic power network simulation algorithms within several products. He was one of the main software developers of RWCap, a parallel program for VLSI capacitance extraction using floating random walk algorithm at Tsinghua University. At UC San Diego, he designed the numerical algorithms for large-scale circuit simulation using matrix exponentials, and optimization algorithms for electrostatics based VLSI global placement. 

Mr. Zhuang was the recipient of the Charles Lee Powell Fellowship and the Qualcomm FMA Fellowship. He is a student member of IEEE, ACM and SIAM. He serves as a reviewer for IEEE Transactions on Computer Aided Design (TCAD), and an external reviewer for Design Automation Conference (DAC) and International Symposium on Physical Design (ISPD). 
\end{IEEEbiography}
\begin{IEEEbiography}[{\includegraphics[width=1in,clip,keepaspectratio]{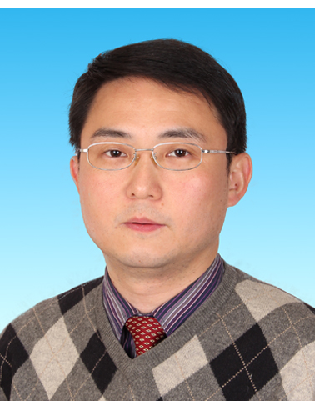}}]
{Wenjian Yu} (S'01-M'04-SM'10) 	received the B.S. and Ph.D. degrees in computer science from Tsinghua University, Beijing, China, in 1999 and 2003, respectively. 
	
In 2003, he joined Tsinghua University, where he is an Associate Professor with the Department of  Computer Science and Technology. He was a Visiting Scholar with the Department of Computer Science  and Engineering, University of California, San Diego, La Jolla, CA, USA, twice during the period from  2005 to 2008. He serves as an associate editor of IEEE Transactions on Computer Aided Design since 2016.
His current research interests include numerical modeling and simulation techniques for  designing complex systems (like integrated circuit, touchscreen, cyber physical system, etc.), and the  matrix computation algorithms for BIG DATA analysis. He has authored three books and over 130 papers in refereed journals and conferences. 
	
Dr. Yu was a recipient of the Distinguished Ph.D. Award from Tsinghua University in 2003 and the Excellent Young Scholar Award from the National Science Foundation of China in 2014.
\end{IEEEbiography} 
\begin{IEEEbiography}[{\includegraphics[width=1in,clip,keepaspectratio]{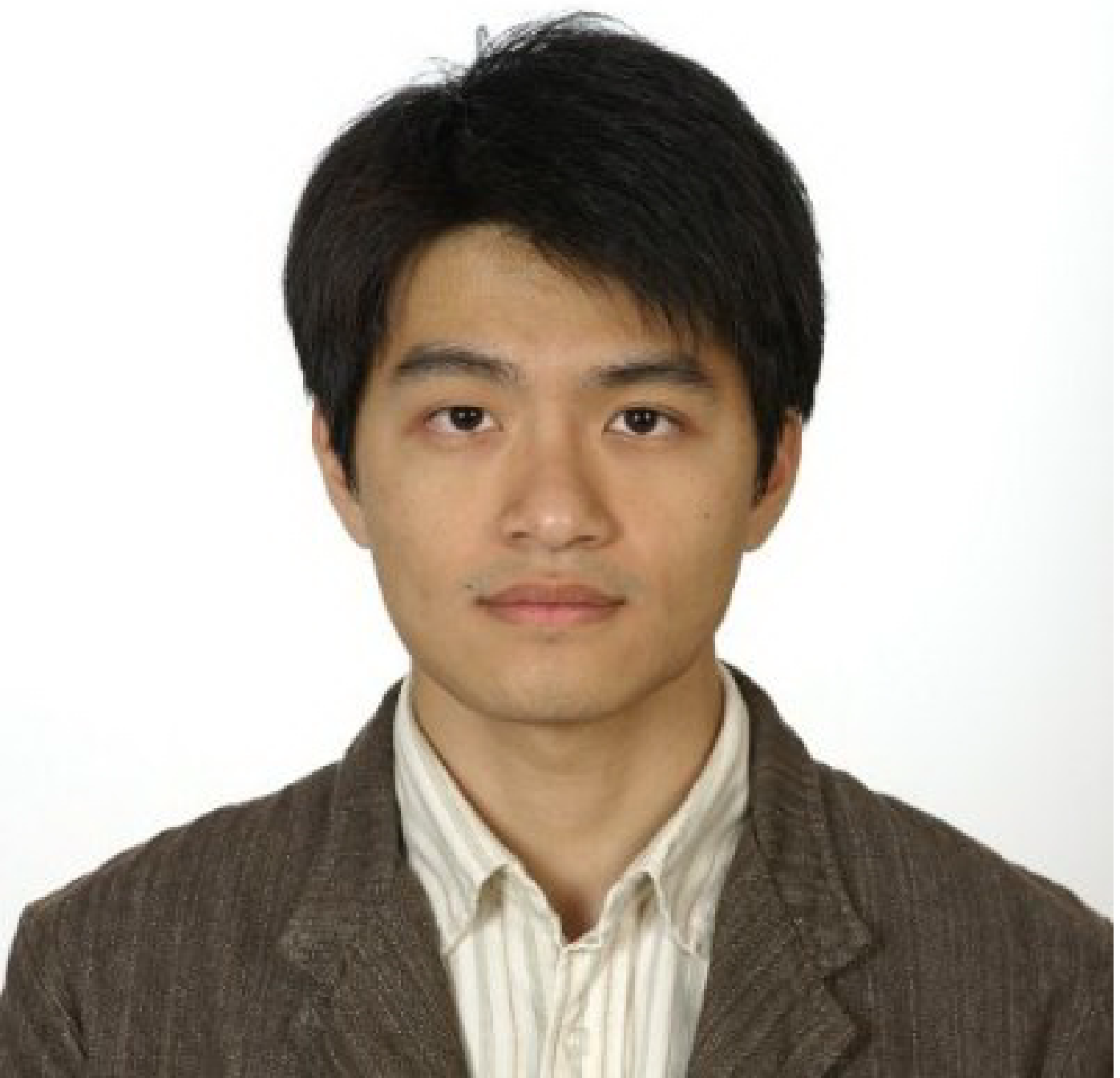}}]{Shih-Hung  Weng} is currently with Facebook as Research Scientist and working on large scale distributed system for payments services. He received the B.S. and M.S degrees in Computer Science from National Tsing Hua University, Hsinchu, Taiwan, in 2006 and 2008, and the Ph.D. degree from University of California San Diego, La Jolla, U.S.A in 2013 under the supervision of Dr. Chung-Kuan Cheng. His thesis focused on parallel circuit simulation for power-grid noise analysis in very large scale integration (VLSI) system. 
\end{IEEEbiography}
\begin{IEEEbiography}[{\includegraphics[width=1in,clip,keepaspectratio]{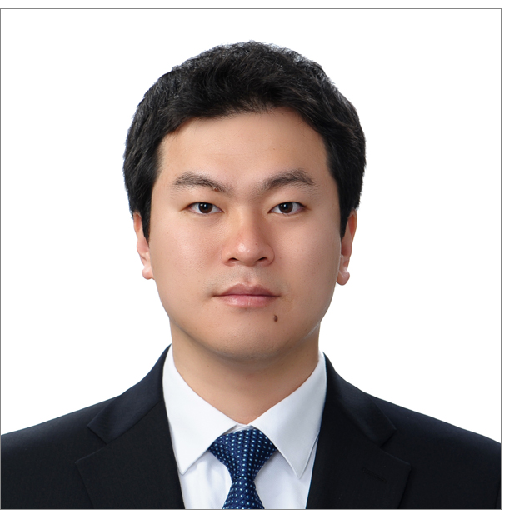}}]{Ilgweon Kang}(S'12) received the B.S and M.S. degrees in electrical and electronic engineering from Yonsei University, Seoul, Korea, in 2006 and 2008, respectively. He is currently pursuing the Ph.D degree at University of California at San Diego (UCSD). He was with the Research and Development Division, SK Hynix, Icheon, Korea, from 2008 to 2012, where he was involved in development of 3D stacked DRAM with (Through Silicon-Via) TSV technology. 
	
His current research interests include 3D IC design and optimization, CAD, and design for testability
\end{IEEEbiography} 
\begin{IEEEbiography}[{\includegraphics[width=1in,height=1.25in,clip,keepaspectratio]{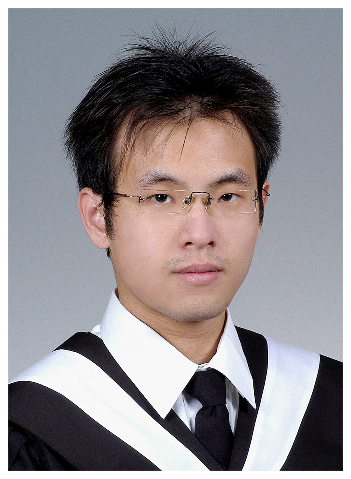}}]{Jeng-Hau Lin} (S'15) is a Ph.D student at Department of Computer Science and Engineering, University of California, San Diego, CA, USA. He received his B.S degree in Electrical Engineering in 2005 and M.S. degree in Communication Engineering in 2007 from National Taiwan University. 
	
	His research interests includes time-frequency analyses on signal integrity and the consensus of multi-agent system.  
\end{IEEEbiography}
\begin{IEEEbiography}[{\includegraphics[width=1in,height=1.25in,clip,keepaspectratio]{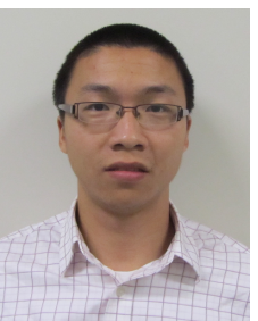}}]{Xiang Zhang} (M'13) received the B.Eng. in Information Engineering from Shanghai Jiao Tong University (SJTU), China and M.S. in 	Electrical and Computer Engineering from University of Arizona, Tucson, AZ, in 2010.  

From 2011 to 2014, he was a Senior Hardware Engineer with Qualcomm Technologies, Inc. In 2014, he joined Apple Inc as an iPhone Hardware System Engineer. Since 2012, he has been working towards the Ph.D. degree in Department of Electrical and Computer Engineering at University of California, San Diego, CA, USA.

His current research interests include power distribution network design and optimization, circuit simulation and system-on-chip design.
\end{IEEEbiography} 
\begin{IEEEbiography}[{\includegraphics[width=1in,height=1.25in,clip,keepaspectratio]{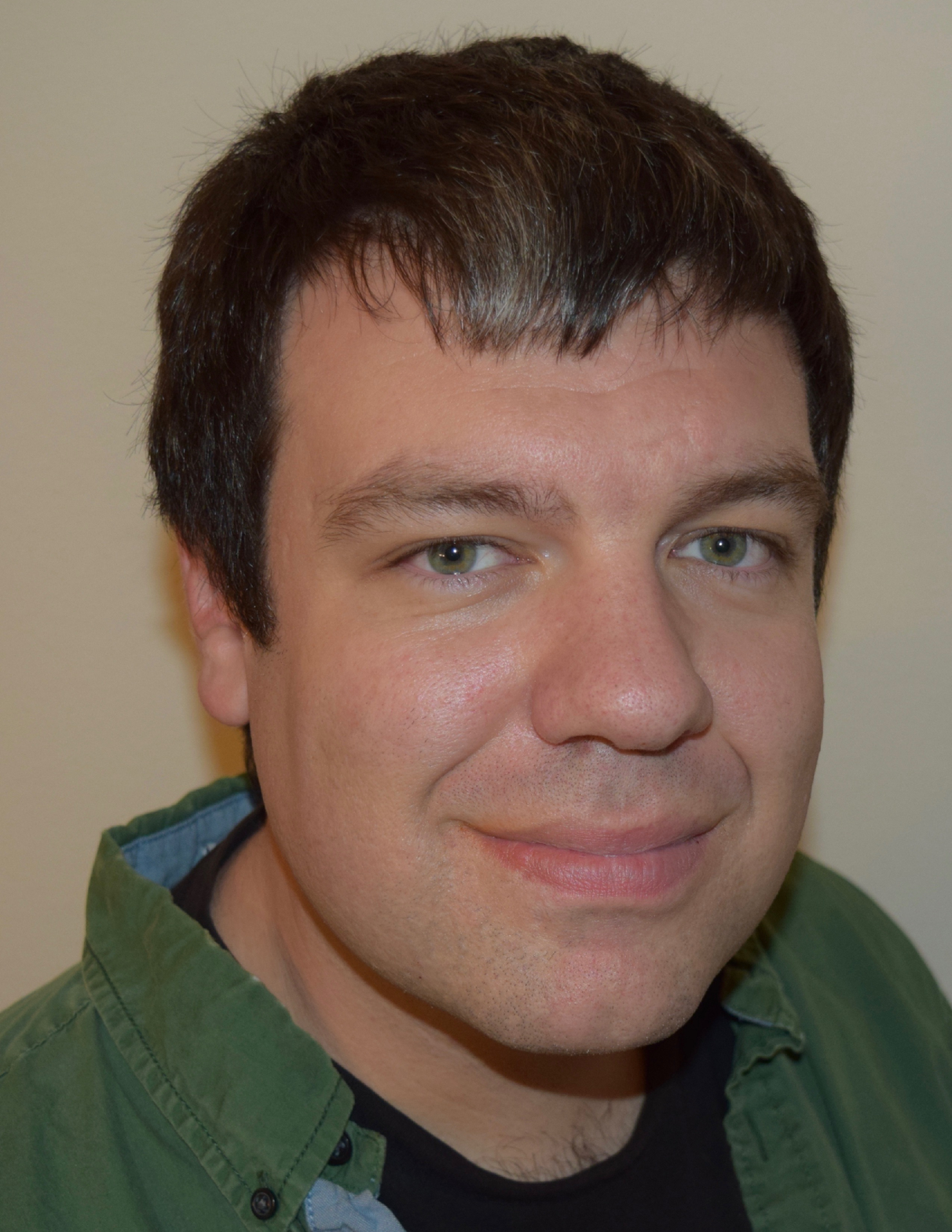}}]{Ryan Coutts}  
received a B.S. in electrical engineering from the University of California, Riverside in 2005.  He then received his M.S. in electrical engineering from Stanford University in 2006. After receiving his M.S. degree, he worked as a signal and power integrity engineer at NVIDIA Inc. specializing in FSB, DDR and power integrity. He is currently working at Qualcomm Inc. where he works in low power design methodologies for mobile covering the range from power integrity, thermal optimization and on-chip timing.  Mr. Coutts has filed 13 patents related to his innovation as well and is currently pursuing a Ph.D at the University of California, San Diego.
\end{IEEEbiography} 
\begin{IEEEbiography}[{\includegraphics[height=1.1in,clip,keepaspectratio]{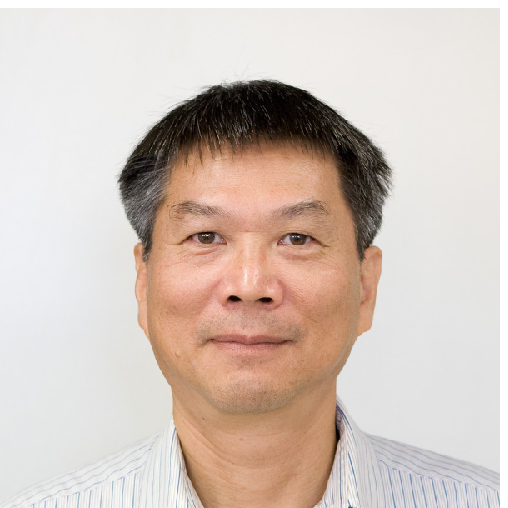}}]{Chung-Kuan Cheng} (S'82-M'84-SM'95-F'00)  
received the B.S. and M.S. degrees in electrical engineering from National Taiwan University, and the Ph.D. degree in electrical engineering and computer sciences
from University of California, Berkeley in 1984.

From 1984 to 1986 he was a senior CAD engineer at Advanced Micro Devices
Inc. In 1986, he joined the University of California, San Diego,
where he is a Distinguished Professor in the Computer Science and
Engineering Department, an Adjunct Professor in the Electrical and
Computer Engineering Department.
He served as a principal engineer at Mentor Graphics in 1999.
He was an associate editor of IEEE Transactions on Computer
Aided Design for 1994-2003.
He is a recipient of the best paper awards, IEEE Trans.
on Computer-Aided Design in 1997, and in 2002, the NCR excellence
in teaching award, School of Engineering, UCSD in 1991,
IBM Faculty Awards in 2004, 2006, and 2007, the Distinguished
Faculty Certificate of Achievement, UJIMA Network, UCSD in 2013.
He is appointed as an Honorary Guest Professor of Tsinghua
University 2002-2008,
and a Visiting Professor of National Taiwan University 2011, and 2015.

His research interests include medical modeling and analysis,
network optimization and design automation on microelectronic circuits.
\end{IEEEbiography}  
 
\end{document}